\documentclass[a4paper,11pt]{article}
\pdfoutput=1
 
\usepackage{jinstpub} 

\usepackage{graphicx}
\usepackage{color}
\usepackage{mathtools}
\usepackage[left]{lineno}
\usepackage{subfig}
\usepackage{hyperref}

\newcommand{\CENBG}{CENBG, Universit\'{e} de Bordeaux, CNRS/IN2P3, 33175 Gradignan, France}
\newcommand{\CEA}{IRFU, CEA, Universit\'{e} Paris-Saclay, F-91191 Gif-sur-Yvette, France}
\newcommand{\CPPM}{ CPPM, Universit\'{e} d'Aix-Marseille, CNRS/IN2P3, F-13288 Marseille, France}
\newcommand{\LSM}{ LSM, CNRS/IN2P3, Universit\'{e} Grenoble-Alpes, Modane, France}
\newcommand{\SUBATECH}{ SUBATECH, IMT-Atlantique, Universit\'{e} de Nantes, CNRS-IN2P3, France}
\newcommand{\LAL}{LAL, Universit\'{e} Paris-Sud, CNRS/IN2P3, Universit\'{e} Paris-Saclay, F-91405 Orsay, France}
\newcommand{\IUF}{Institut Universitaire de France, F-75005 Paris, France}

\title{Study of a spherical Xenon gas TPC for neutrinoless double beta detection}

\author[a]{A.~Meregaglia,}
\author[b]{J.~Busto,}
\author[a]{C.~Cerna,}
\author[a]{M.~Chauveau,}
\author[c]{A.~Dastgheibi-Fard,}
\author[a]{C.~Jollet,}
\author[f]{S.~Jullian,}
\author[d]{I.~Katsioulas,}
\author[d]{I.~Giomataris,}
\author[d]{M.~Gros,}
\author[e]{P.~Lautridou,}
\author[a]{C.~Marquet,}
\author[d]{X.~F.~Navick,}
\author[a]{F.~Perrot,}
\author[a,c]{F.~Piquemal,}
\author[f,g]{L.~Simard,}
\author[c]{M.~Zampaolo}

\affiliation[a]{\CENBG}
\affiliation[b]{\CPPM}
\affiliation[c]{\LSM}
\affiliation[d]{\CEA}
\affiliation[e]{\SUBATECH}
\affiliation[f]{\LAL}
\affiliation[g]{\IUF}

\emailAdd{amerega@in2p3.fr}

\abstract{
 Several efforts are ongoing for the development of spherical gaseous time projection chamber detectors for the observation of rare phenomena such as weakly interacting massive particles or neutrino interactions. The proposed detector, thanks to its simplicity, low energy threshold and energy resolution, could be used to observe the $\beta\beta0\nu$ process i.e. the neutrinoless double beta decay. In this work, a specific setup is presented for the measurement of $\beta\beta0\nu$ on 50~kg of $^{136}$Xe. The different backgrounds are studied, demonstrating the possibility to reach a total background per year in the detector mass at the level of 2 events per year. The obtained results are competitive with the present generation of experiments and could represent the first step of a more ambitious roadmap including the $\beta\beta0\nu$ search with different gases with the same detector and therefore the same background sources. The constraints in terms of detector constructions and material purity are also addressed, showing that none of them represents a show stopper for the proposed experimental setup.}

\begin{document} 

\maketitle
\flushbottom


\section{Introduction}
In the last decade significative efforts were made in the development of spherical gaseous time projection chamber (TPC) detectors for the observation of low energy weakly interacting massive particles (WIMPs) as dark matter candidate.\\
The main motivation for such a detector relies on its simplicity given by a single channel readout, the low energy threshold as small as 17~eV allowing to observe a single electron signal, and the good extrapolated energy resolution at the level of 0.4\% RMS at 2.4~MeV~\cite{Gerbier:2014jwa,Arnaud:2017bjh}.\\
Recently the possibility to use such a detector for the search of neutrinoless double beta decays ($\beta\beta0\nu$) has been considered~\cite{gdr} using $^{136}$Xe gas to fill the TPC sphere. The use of Xenon is motivated  by its natural abundance of $^{136}$Xe at $\sim 9\%$, its low cost and the fact that it is safe and easy to enrich.\\
In this paper a possible setup of the detector is studied in detail, showing that its performance in terms of background events in the region of interest (ROI) can be as low as the one reached by other Xenon based experiments such as EXO~\cite{Auger:2012ar}, NEXT~\cite{Lorca:2012dv} or KamLAND-Zen~\cite{Gando:2012zm}.\\
The possibility to set a limit on the $\beta\beta0\nu$ decay using $^{136}$Xe in a moderate volume detector would definitely be an important achievement, although it would be difficult to obtain competitive results compared to already running experiments with larger mass. The real breakthrough would be the possibility to use the same detector filled with different gases and to study the $\beta\beta0\nu$ process for different isotopes (possibly with higher $Q_{\beta\beta}$ than $^{136}$Xe for an easier background rejection) and the same background sources.

\section{Detector and simulation setup}
\label{sec:simu}
\begin{figure} [t]
\begin{center}
\includegraphics[width=0.45\textwidth]{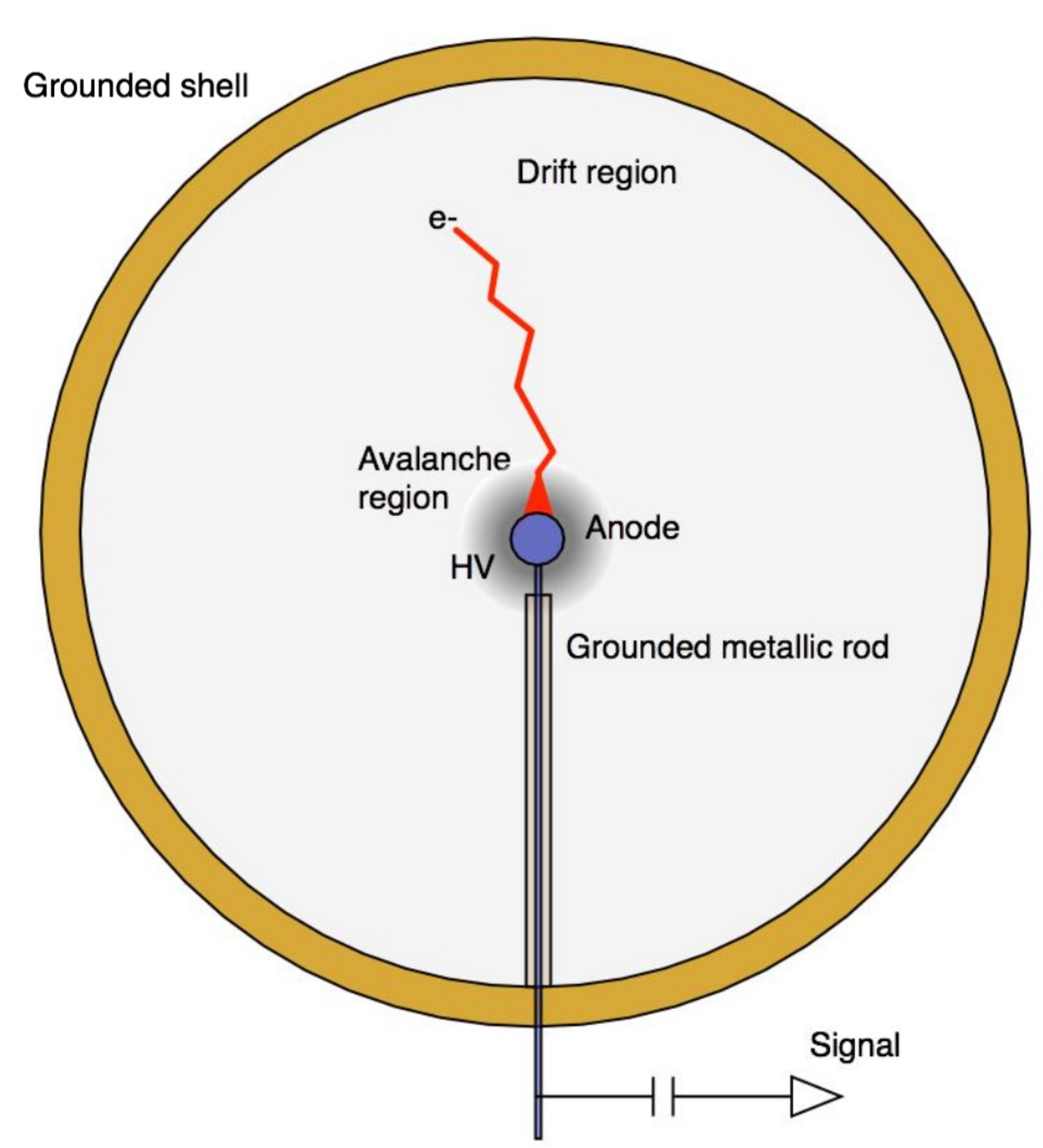}
\caption{{\it Working principle of the sphere TPC.}}
\label{fig:TPC}
\end{center}
\end{figure}

\begin{figure} [t]
\begin{center}
\includegraphics[width=0.95\textwidth]{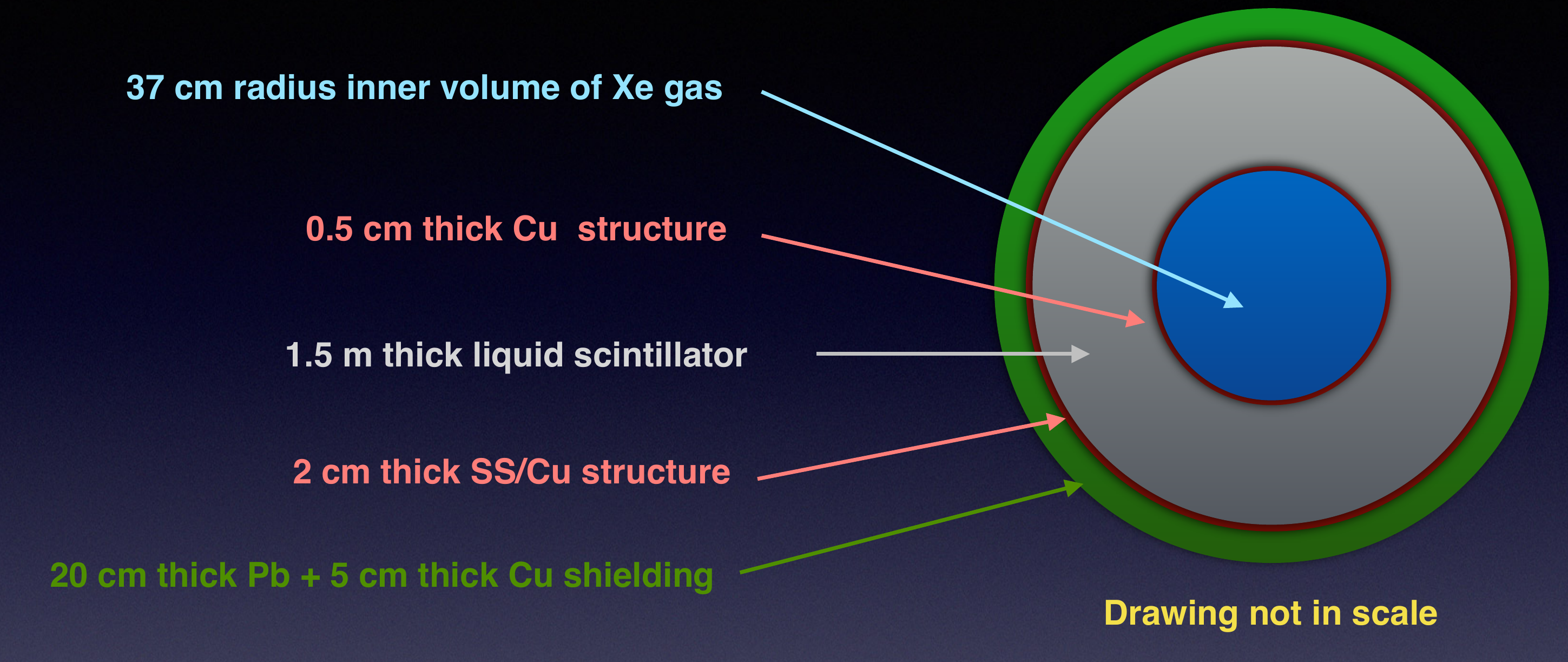}
\caption{{\it Schematic view of the detector setup.}}
\label{fig:Setup}
\end{center}
\end{figure}

The working principle of the sphere TPC can be seen in Fig.~\ref{fig:TPC}: the electrons produced by the particles crossing and ionizing the Xenon gas are drifted towards the central anode. When the electrons are close enough to the anode they enter the avalanche region and they are collected by the anode itself to form the readout signal (see Ref.~\cite{Gerbier:2014jwa,Savvidis:2016wei} for more details).\\ 
In the GEANT4~\cite{Agostinelli:2002hh} based simulation developed for the presented studies, the electron drift and their avalanche were not considered: all the results are based on the Monte-Carlo (MC) information on the true energy deposition of the different particles  where no quenching for $\alpha$'s is accounted for. This simplification does not affect the robustness of the results since no signal selection was made based on the reconstructed waveform.
A Gaussian energy smearing on the deposited energy is applied before any selection cut: an energy resolution of 1\% FWHM was assumed at $^{136}$Xe $Q_{\beta\beta}$ of 2.458~MeV~\cite{Redshaw:2007un} with an energy dependence proportional to $\sqrt{E}$ i.e. $\frac{\Delta E_{FWHM}}{E}= \frac{1\%}{\sqrt{E/Q_{\beta\beta}}}$.\\
The detector geometry was defined as follows (see schematic view in Fig.~\ref{fig:Setup}):
\begin{itemize}
\item {\it Xenon active volume}\\ Inner sphere of $^{136}$Xe active volume with a radius of 37~cm and a gas pressure of 40~bar  corresponding to a total mass of about 50~kg. This choice, based on the results of a pressure and radius scan, is driven by the need of containing at least 80\% of the $\beta\beta0\nu$ electrons. Unless otherwise stated (e.g. in the neutron capture studies on the different Xenon isotopes of Sec.~\ref{sec:NeutronCapture}), an enriched gas of pure $^{136}$Xe is considered.\\
\item {\it Xenon vessel}\\ The active volume is contained in a Copper sphere with a thickness of 0.5~cm. Preliminary  studies show that such a thickness is enough to assure the mechanical stability of the detector, and of course the thinner the Copper the smaller the background expected in the active volume.\\
\item {\it Liquid scintillator volume}\\ The Copper sphere is contained in a spherical liquid scintillator volume, assumed to be Linear Alkylbenzene (LAB) in our MC, with a thickness of 1.5~m (i.e. a radius of 1.875~m) which is used as veto volume. Such a veto provides a rejection of both events coming from the outside, and radioactive events generated in Copper, in particular $^{208}$Tl as reported in Sec.~\ref{sec:BGrejection}.\\
\item {\it Liquid scintillator vessel}\\ The vessel is a sphere of a thickness of 2~cm. Stainless steel was considered as a first option, however, given the results obtained on the background contribution due to the vessel radioactivity (see Sec.~\ref{sec:LSBG}) and on the trigger rate in the liquid scintillator (see Sec.~\ref{sec:LSrate}), a vessel made of Copper seems a most appropriate choice.
\item {\it Shielding}\\ An external sphere of 20~cm of Lead and 5~cm of Copper was assumed as shielding. A smaller thickness resulted in a too large background rate from external gammas as reported in Sec.~\ref{sec:ExtGamma}. The choice of using Lead and Copper is due to the fact that the study of external gammas is based on measurements performed at the Underground Modane Laboratory (LSM) which used a shielding made of Lead and Copper therefore the use of the same shielding resulted in a more reliable and less complicated MC.
\end{itemize}
In the following studies both the possible signal ($\beta\beta2\nu$ and $\beta\beta0\nu$) and different backgrounds were simulated. Since GEANT4 does not include yet the double beta decay process for $^{136}$Xe, the energy of the two electrons, both in the 2$\nu$ and in the 0$\nu$ case, is obtained from pre-computed spectra~\cite{BBspectra,Kotila:2012zza}. Their angular correlation has not been included so far but the impact is expected to be a second order effect.\\
For what concerns the backgrounds, $^{60}$Co decays, $^{232}$Th and  $^{238}$U decay chains were simulated. A conservative time window of 10~ms was assumed to separate events in the decay chain considering the typical drift time of a few tens of $\mu$s and the signal width of the order of 1~ms. Given the half-life of 164~$\mu$s for $^{214}$Po ($^{238}$U chain) and 300~ns for $^{212}$Po ($^{232}$Th chain), Bi-Po decays are seen as a single event in our detector simulation.


\section{Signal selection cuts}
\label{sec:BGandROI}
In this section the background rejection is studied considering the radioactive decays produced in the Copper vessel containing the active Xenon volume. The different decay chains are simulated inside the Copper volume with a constant probability per unit volume. Based on the obtained results the ROI is further optimised to maximise the possible limit which could be set on the half-life $T_{1/2}^{0\nu}$ for the $\beta\beta0\nu$ decay. The obtained ROI and the applied strategy for the background rejection will be used unchanged in the following sections for the studies of all the possible additional sources of background.\\

\begin{figure}[t]
\begin{center}
\includegraphics[width=0.95\textwidth]{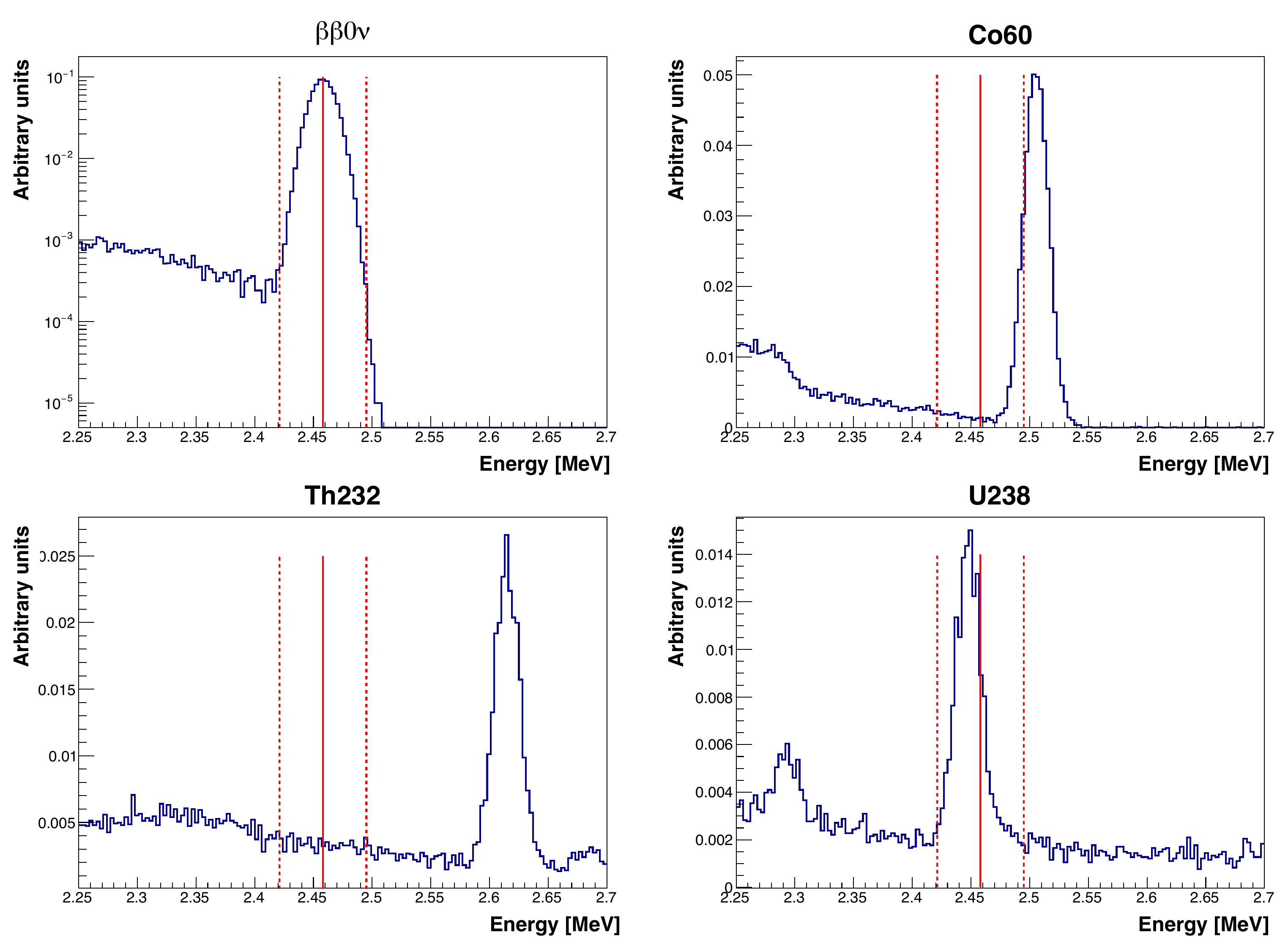}
\caption{{\it Energy spectrum in arbitrary units (normalized to 1) of expected $\beta\beta0\nu$ signal (top left), $^{60}$Co background (top right),  $^{232}$Th chain background (bottom left) and $^{238}$U chain background (bottom right). The $Q_{\beta\beta}$ is shown by the solid red line whereas the limits of ROI of $\pm 1.5\%$ with respect to the $Q_{\beta\beta}$ are shown by the dashed red lines.}}
\label{fig:ene}
\end{center}
\end{figure}

\subsection{Background rejection strategy}
\label{sec:BGrejection}

\begin{figure}[t]
\begin{center}
\includegraphics[width=0.7\textwidth]{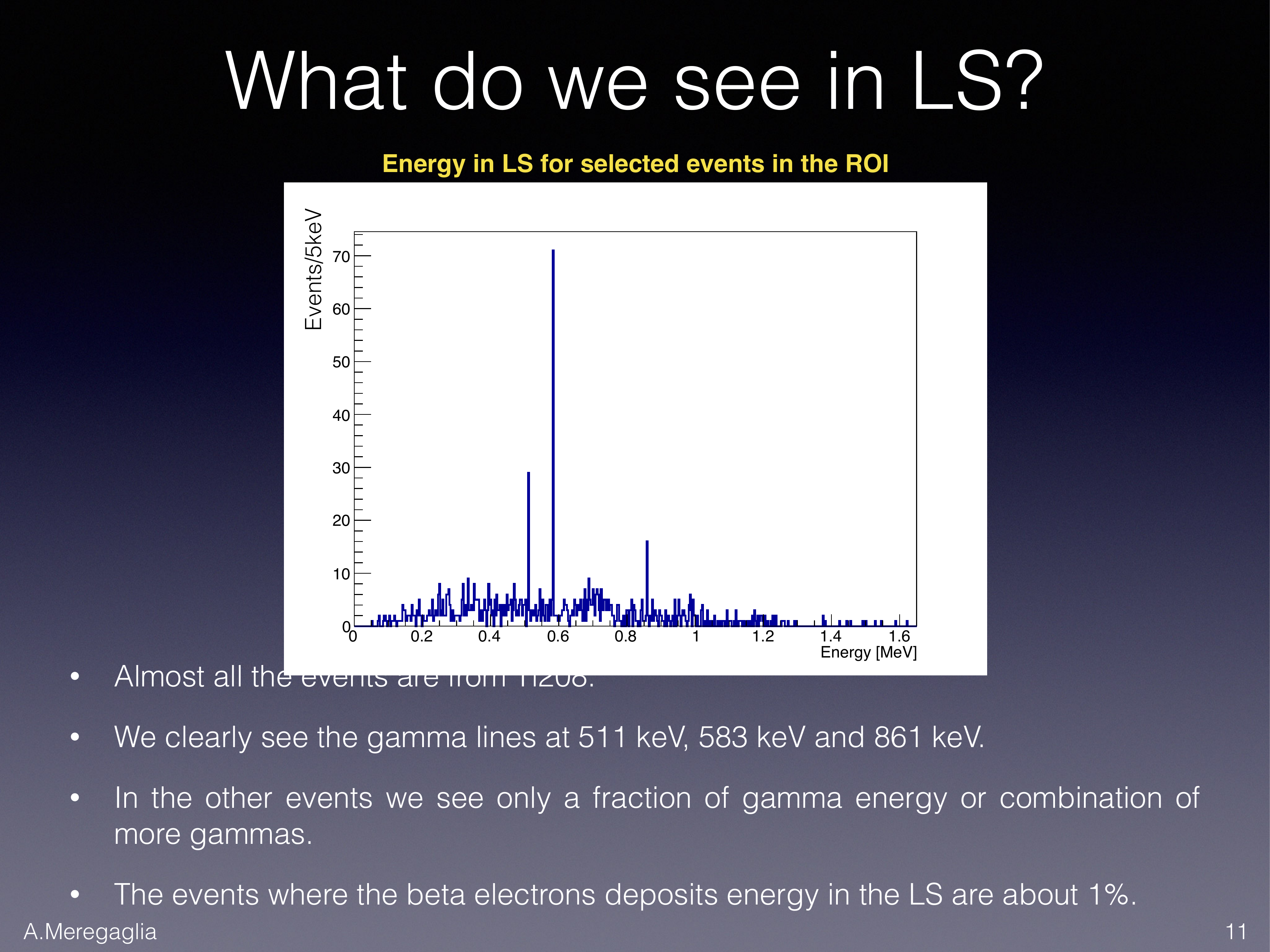}
\caption{{\it Energy deposited in the liquid scintillator for events of the $^{232}$Th decay chain selected in the ROI according to their energy deposition in the Xenon active volume.}}
\label{fig:LS}
\end{center}
\end{figure}

\begin{figure}[tbph]
\begin{center}
\includegraphics[width=0.95\textwidth]{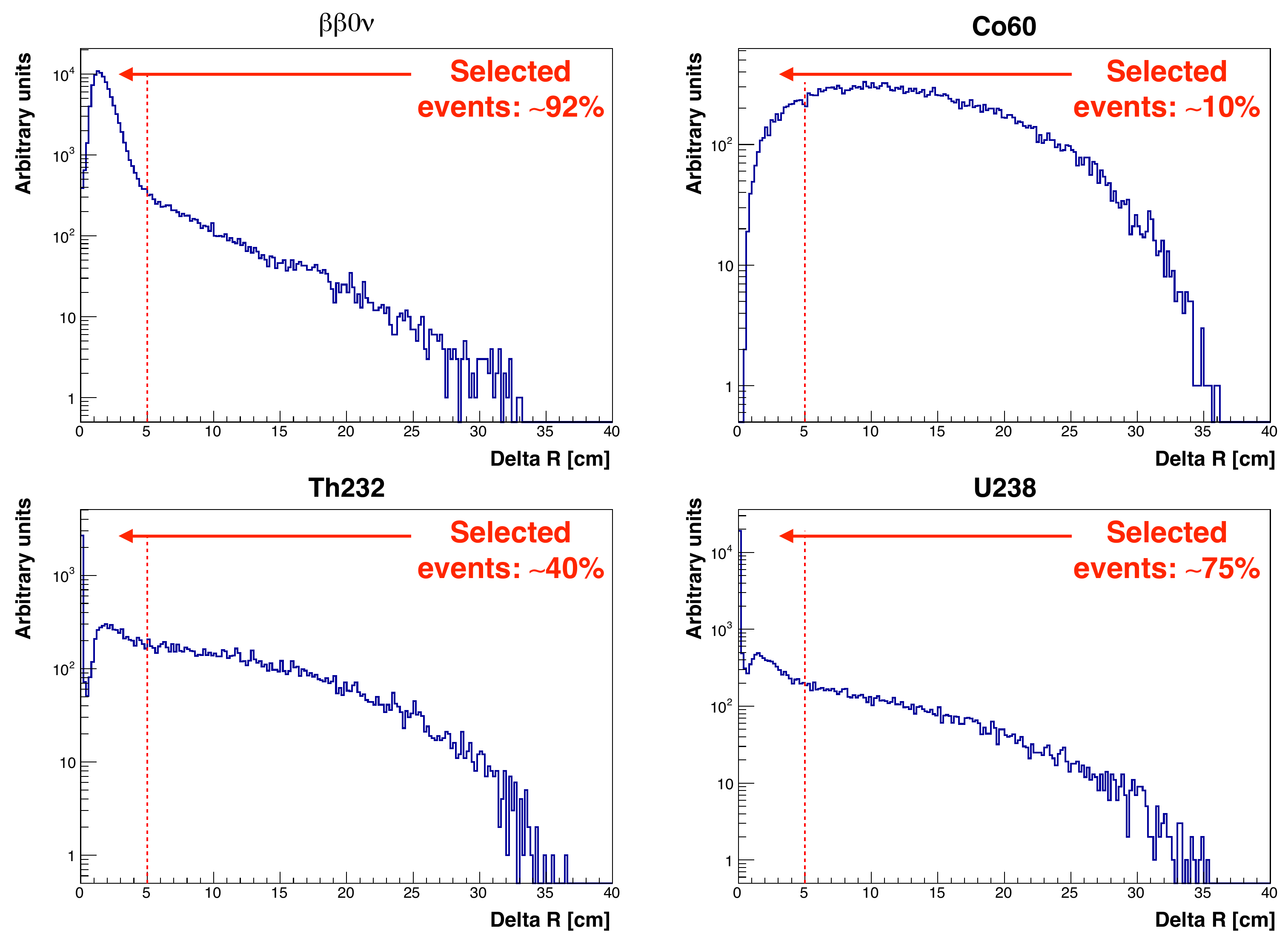}
\caption{{\it $\Delta$R distribution for $\beta\beta0\nu$ signal (top left), $^{60}$Co background (top right),  $^{232}$Th chain background (bottom left) and $^{238}$U chain background (bottom right) before ROI and LS based selection. The selected region with $\Delta$R $< 5$~cm is shown by the dashed red line.}}
\label{fig:DR}
\end{center}
\end{figure}

The first step in the background rejection is given by the ROI selection. The different expected energy spectra for the signal and for the simulated background can be seen in Fig.~\ref{fig:ene}, where a ROI of $\pm 1.5\%$ with respect to the $Q_{\beta\beta}$ is also shown. With the excellent energy resolution, assumed in the Gaussian smearing of the deposited energy as explained before, we see that the $^{208}$Tl line at 2.6~MeV, issued by the $^{232}$Th decay chain, is clearly out of the ROI whereas a tail of the 2.5~MeV of the two gammas from $^{60}$Co and a peak due to the $^{214}$Bi decay in the Uranium decay chain are inside the ROI.\\
A reduction of the background can be achieved exploiting the liquid scintillator (LS): events with an energy deposition in the liquid scintillator volume larger than 200~keV are rejected. Such a selection is particularly effective for $^{208}$Tl events (i.e. almost the totality of the background issued by the $^{232}$Th decay chain) where gammas emitted in coincidence with the 2.6~MeV gamma line (possible source of signal in the ROI when not all the energy is deposited inside the Xenon active volume) can be identified. The energy deposited in the liquid scintillator for $^{232}$Th events in the ROI can be seen in Fig.~\ref{fig:LS}: almost all the events are from $^{208}$Tl and we clearly see the gamma lines emitted at 511~keV, 583~keV and 861~keV as expected~\cite{NuclearData}. The other events correspond to only a fraction of the energy of a gamma, or to a combination of more gammas. The fraction of events in which the $\beta$ electron is observed in the liquid scintillator corresponds to about 1\% of the observed energy depositions, since the electrons lose almost all their energy in the Copper volume. Test with a lower threshold in the LS at the level of 50~keV were performed showing an almost negligible improvement in the background rejection with the drawback of a more complicated light readout.\\
An additional handle for a further background reduction can be given by the radial energy deposition distribution in the Xenon volume. The feasibility studies of such a radial reconstruction, based on the width of the measured signal affected by drifted electron diffusion, are ongoing. The preliminary results obtained with small prototypes at LSM are promising, however, in case the obtained precision is not sufficient, the possibility to read out the Xenon scintillation light and use it as a trigger for the drift starting time can be considered. 
Looking at the maximal radial distance between the energy deposition points (so called $\Delta$R distribution) for the signal and the different background (see Fig.~\ref{fig:DR}) it can be seen that about 92\% of the events have a $\Delta$R smaller than 5~cm. 
With such a selection cut, a large fraction of background could be rejected, in particular considering the $^{60}$Co since the energy depositions of the two gammas at different radial positions result into a rejection of $\sim 90\%$ of the events.\\
It can also be noticed that $^{232}$Th and $^{238}$U distributions show a peak at very small (i.e. almost 0) $\Delta$R: this is due to $\alpha$ events releasing all their energy very close to the Copper surface. Therefore, an additional cut based on the fiducial volume (so called Rmin cut), can be applied to reject events for which the minimal radial distance is larger than 36~cm (i.e. 1 cm from the Copper surface).\\
\begin{table}[t]
\begin{center}
\begin{tabular}{|c|c|c|c|c|}
\hline
Source & Events in ROI & + LS cut & + Rmin cut & + $\Delta$R cut\\
 & ($Q_{\beta\beta} \pm 1.5\%$) & (Energy in LS$<200$~keV)& (Rmin$<36$~cm) & ($\Delta$R$< 5$~cm) \\

\hline
$^{60}$Co & $6.9 \pm 0.1$ & $6.9 \pm 0.1$ & $6.9 \pm 0.1$ & $0.8 \pm 0.1$\\
$^{232}$Th chain& $28.0 \pm 0.8$ & $4.6 \pm 0.3$ & $3.7 \pm 0.3$ & $0.8 \pm 0.1$\\
$^{238}$U chain & $9.3 \pm 0.2$ & $8.8 \pm 0.2$ & $7.5 \pm 0.2$ & $3.0 \pm 0.1$\\
\hline
$\beta\beta0\nu$ & $82.1\% \pm 0.3\%$ & $82.1\% \pm 0.3\%$ & $81.9\% \pm 0.3\%$ & $76.2\% \pm 0.3\%$\\
\hline
\end{tabular}
\end{center}
\caption{{\it Number of background events per year for different background sources as a function of the cuts which are applied on top of each other from left to right. The $\beta\beta0\nu$ signal efficiency is also reported.}}
\label{tab:BG}
\end{table}%
Assuming a Copper activity of 10~$\mu$Bq/kg in each source of background (i.e. each element of the decay chain for $^{232}$Th and $^{238}$U), the number of expected background events per year was computed for the considered setup, reaching a total rate of about 5 events after all cuts, as can be seen in Tab.~\ref{tab:BG}.\\
Considering that secular equilibrium in the Uranium chain can be assumed only starting from $^{226}$Ra, the results were cross-checked simulating the chain starting indeed from Radium. The outcomes ($3.0 \pm 0.1$ for $^{238}$U and $3.1 \pm 0.1$ for $^{226}$Ra) confirm that the simulation of the full Uranium chain does not bias the final background estimate.\\

\subsection{ROI optimization}
\begin{figure}[t]
\begin{center}
\includegraphics[width=0.99\textwidth]{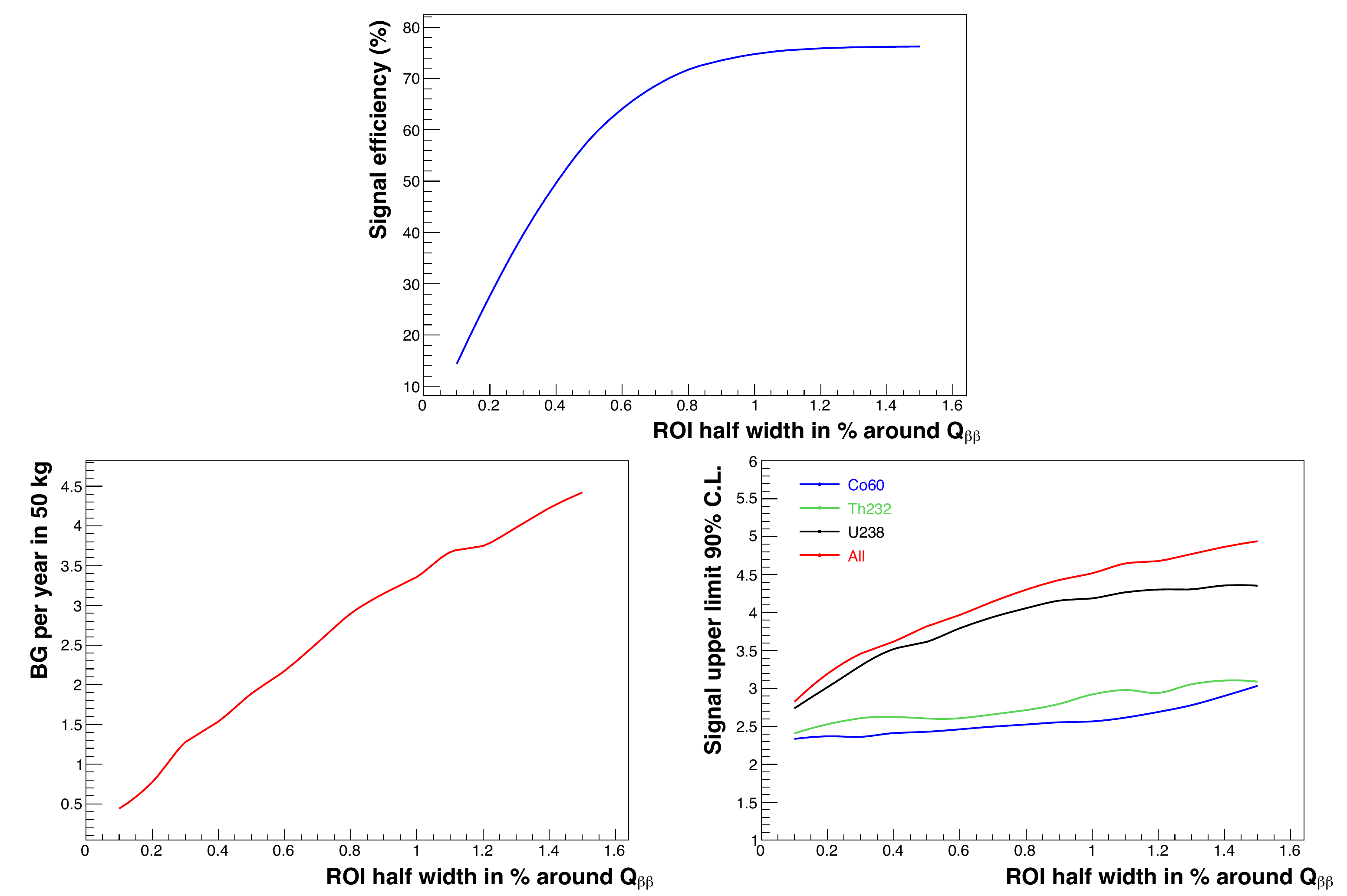}
\caption{{\it ROI scan centred on $Q_{\beta\beta}$: signal detection efficiency (top), background events per year (bottom left) and signal upper limit (bottom right).}}
\label{fig:ROIscan}
\end{center}
\end{figure}

\begin{figure}[t]
\begin{center}
\includegraphics[width=0.7\textwidth]{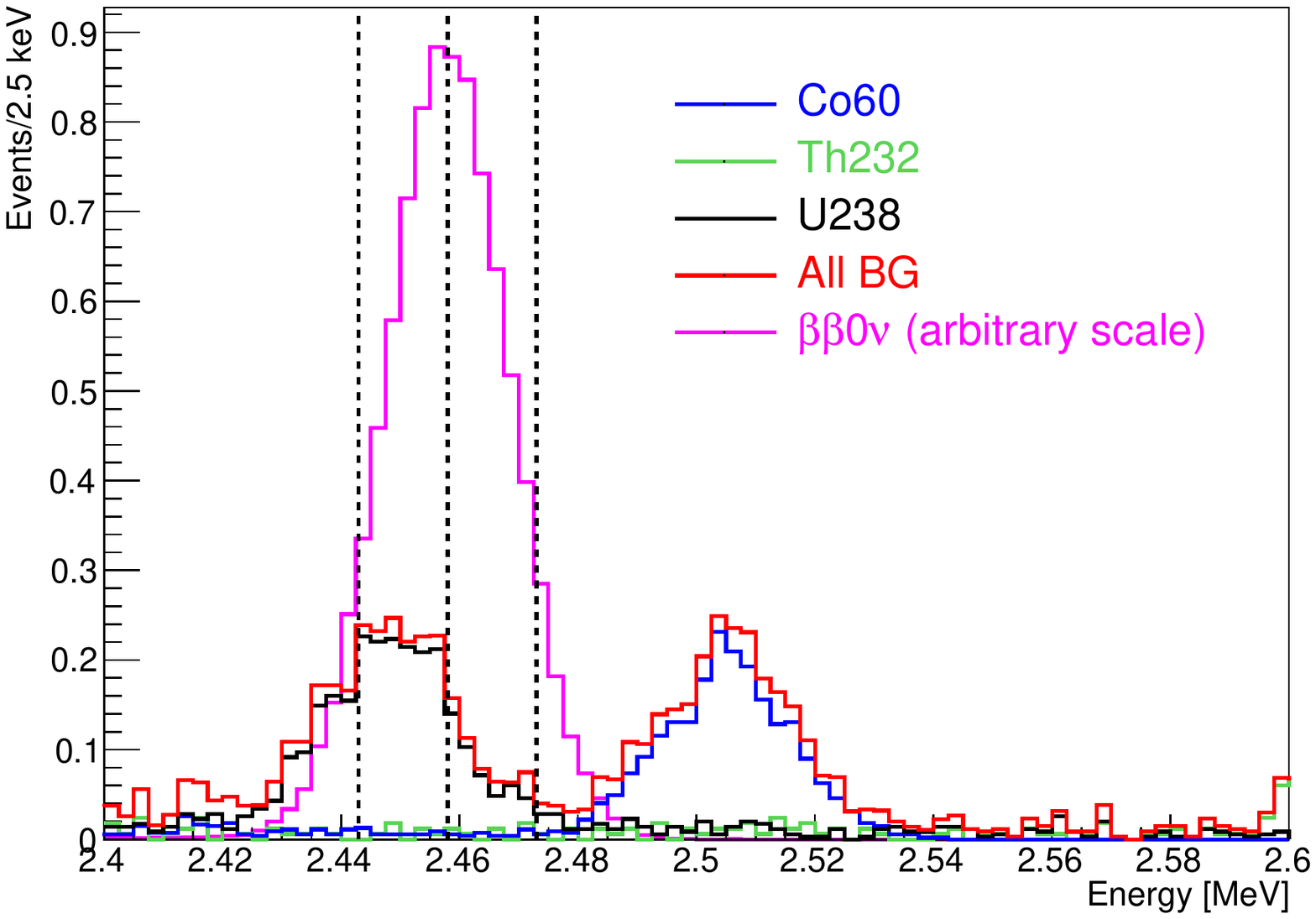}
\caption{{\it Background events after the selection cuts in the 50~kg setup considering 1 year of data taking. The vertical lines represent the ROI at $\pm 0.6\%$ of the $Q_{\beta\beta}$.}}
\label{fig:summary}
\end{center}
\end{figure}

For the study of the background reduction cuts a ROI around the Xenon $Q_{\beta\beta}$ with  a width of 1.5\% was assumed. However, the definition of the ROI can be optimized to maximise the experimental outreach. The variation of several parameters as a function of the ROI were studied, although the final choice of the ROI relies on the maximisation of the limit which could be set on the half-life $T_{1/2}^{0\nu}$.\\
The parameters considered are:
\begin{itemize}
\item Total number of background events per year in the 50~kg Xenon active volume.\\
\item Signal detection efficiency in the 50~kg Xenon active volume.\\ 
\item Signal upper limit ($S_{up}$).\\ It is computed using a Bayesian upper limit for a Poisson parameter, using a uniform prior p.d.f., as explained in Ref.~\cite{Agashe:2014kda}:
\begin{equation}
S_{up}=\frac{1}{2} F^{-1}_{\chi^2} [p,2(n+1)]-b
\end{equation}
where $F^{-1}_{\chi^2}$ is the quantile of the $\chi2$ distribution (inverse of the cumulative distribution), $b$ the expected background and $n$ the number of observed events. The quantity $p$ is defined as:
\begin{equation}
p=1-\alpha(F_{\chi^2}[2b,2(n+1)])
\end{equation}
where $F_{\chi^2}$ is the cumulative $\chi^2$ distribution and $(1-\alpha)$ the confidence level.
\item Limit on $T_{1/2}^{0\nu}$.\\
It is computed according to the following formula:
\begin{equation}
T_{1/2}^{0\nu}>ln(2) \cdot \frac{\epsilon}{S_{up}} \cdot \frac{N_A m}{M} \cdot  T
\end{equation}
where $\epsilon$ is the signal efficiency, ${N_A}$ the Avogadro's number, ${M}$ the Xenon molar mass in g, $T$ the exposure time in years, $m$ the Xenon mass in g and $S_{up}$ the signal upper limit computed before for a C.L. of 90\%.\\
\end{itemize}
\begin{figure}[t]
\begin{center}
\includegraphics[width=0.7\textwidth]{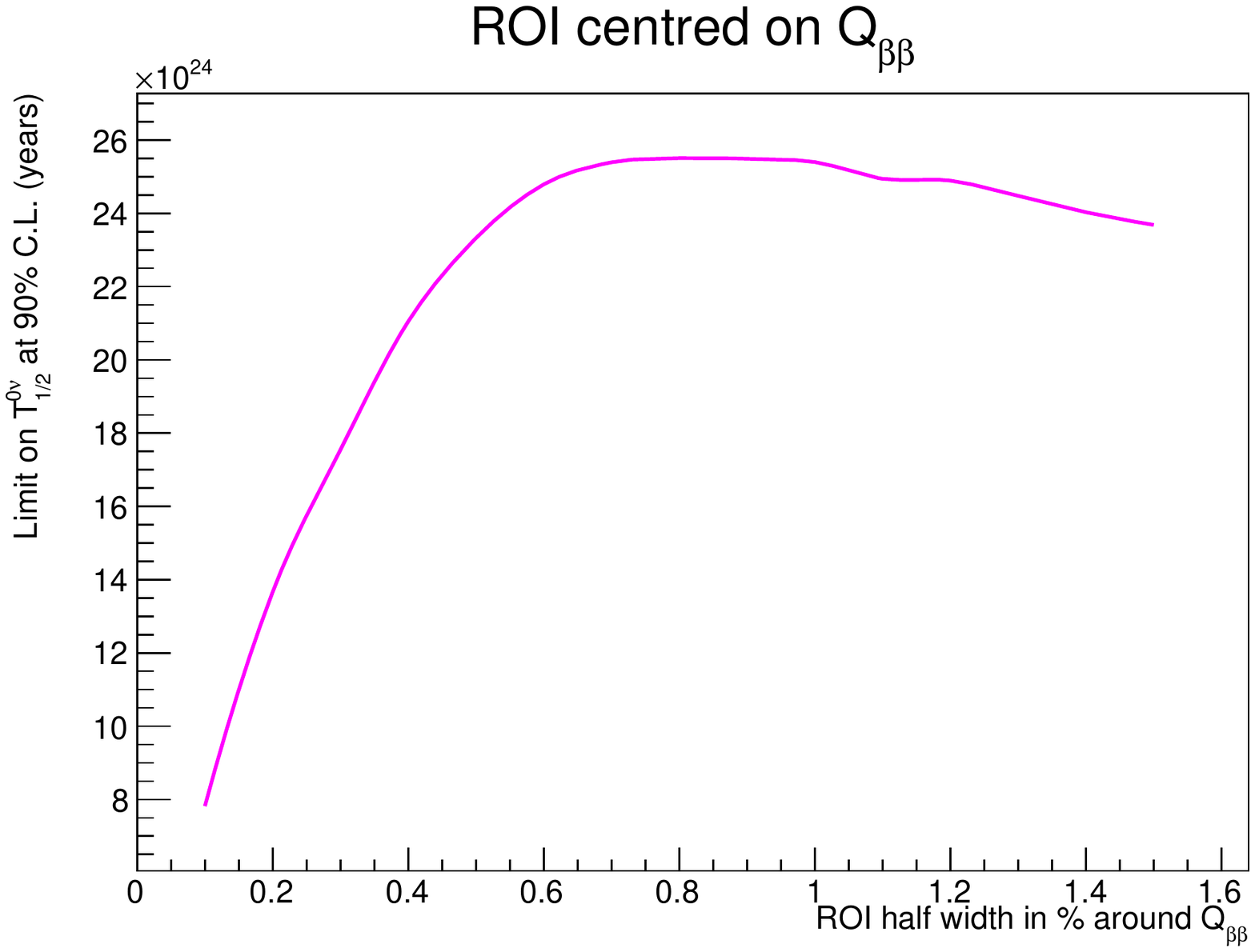}
\caption{{\it $T_{1/2}^{0\nu}$ limit for ROI centred on $Q_{\beta\beta}$ as a function of the ROI half width in \%.}}
\label{fig:TLimit}
\end{center}
\end{figure}
Different ROI scans were performed namely:
\begin{itemize}
\item ROI centred on the $Q_{\beta\beta}$ with varying width (up to $\pm 1.5\%$).
\item ROI starting at  $Q_{\beta\beta} - 1.5\%$ with increasing width up to  $Q_{\beta\beta} + 1.5\%$.
\item ROI starting at $Q_{\beta\beta} + 1.5\%$ with decreasing width down to $Q_{\beta\beta} - 1.5\%$.
\item ROI of 5~keV scanning the region between $Q_{\beta\beta} - 1.5\%$ and $Q_{\beta\beta} + 1.5\%$.
\end{itemize}
The best performance can be achieved for a ROI centred on $Q_{\beta\beta}$ and the largest sensitivity of $T_{1/2}^{0\nu} > 2.5 \times 10^{25}$ years is obtained with a symmetric width larger than $\pm 0.6\%$ i.e. an energy window of [2.443--2.473]~MeV, as can be seen in Fig.~\ref{fig:TLimit}, whereas all the other studied parameters are shown in Fig.~\ref{fig:ROIscan}. Larger ROI up to 1.2\% width would result into a very similar sensitivity but a larger background level.\\
Using the optimal ROI selected based on the $T_{1/2}^{0\nu}$ limit, the expected background amounts to $\sim 2$ events per year whereas the signal upper limit is $\sim 4$ events at a C.L. of 90\% and the signal efficiency is $\sim 64\%$. In Fig.~\ref{fig:summary} a summary picture of the different background components is shown. Note that almost the totality of the remaining background is due to the $^{238}$U chain: the detailed breakdown is 0.1 events per year from $^{60}$Co, 0.3 events per year from $^{232}$Th and 1.9 events per year from $^{238}$U.\\
The definition of the ROI as $Q_{\beta\beta} \pm 0.6\%$ will be used in the following sections when addressing the possible additional sources of background.\\

\section{Background studies}
\subsection{$\beta\beta2\nu$}
The signal issued by the $\beta\beta2\nu$ process could be a source of background for events in the tail region.\\
Assuming in a conservative way  a 50~kg mass of pure $^{136}$Xe and a mean lifetime $T_{1/2}^{2\nu}$ of $2.165 \times 10^{21}$ years~\cite{Agashe:2014kda}, the total number of decays expected per year is  about 70000. Considering the energy resolution of 1\% FWHM at $Q_{\beta\beta}$, the fraction of events in the ROI is $\sim 10^{-9}$ which corresponds to less than $10^{-4}$ events per year.\\
The $\beta\beta2\nu$ contribution to the background is therefore completely negligible.

\subsection{Background from liquid scintillator vessel}
\label{sec:LSBG}
\begin{figure}[t]
\begin{center}
\includegraphics[width=0.7\textwidth]{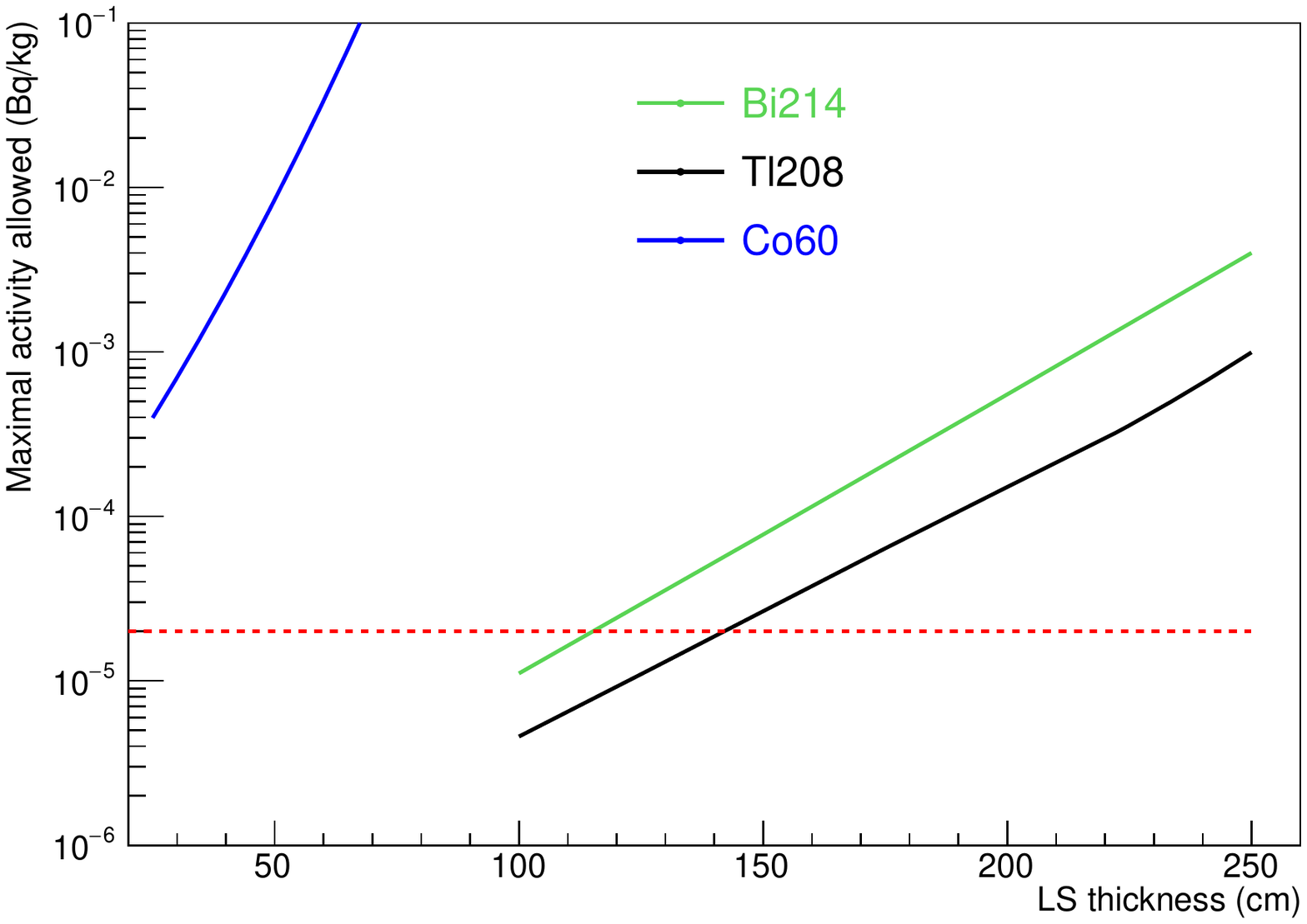}
\caption{{\it Maximal allowed activity of $^{214}$Bi (green), $^{208}$Tl (black) and $^{60}$Co (blue) in Bq/kg as a function of the liquid scintillator thickness to have 0.1 background events (after all cuts) per year in the 50~kg Xenon active volume. The dashed red line represents a limit of 20~$\mu$Bk/kg.}}
\label{fig:extSS}
\end{center}
\end{figure}

The $^{60}$Co,  $^{208}$Tl and $^{214}$Bi contamination of the material used as vessel for the liquid scintillator could result in a source of background.\\
The first natural choice for the 2~cm thick spherical vessel with an inner radius of 187.5~cm was stainless Steel. A sample of $10^{9}$ events for each background was generated uniformly in the Steel volume, and the number of events passing the selection cuts as explained in Sec.~\ref{sec:BGandROI} was computed. Using the obtained rejection fraction, and asking for such a background to be negligible (i.e. smaller than 0.1 events per year), the maximal allowed activity of the Steel can be computed. The activity was indeed computed for different values of the liquid scintillator thickness and the results are shown in Fig.~\ref{fig:extSS}, where the increase of stainless Steel mass at larger liquid scintillator thickness (i.e. larger stainless steel sphere radius) is already accounted for.\\
It is clear that the Cobalt contamination does not represent an issue: for a liquid scintillator thickness of 1.5~m, as assumed in the studied setup, activities higher than 1~Bq/kg can be accepted, which is a limit much higher than what can be expected. On the contrary, the $^{208}$Tl contamination is critical: for a liquid scintillator thickness of 1.5~m, an activity smaller than about 25~$\mu$Bq/kg is required, whereas the typical activity in stainless Steel is of the order of a few mBq/kg~\cite{Maneschg:2008zz} i.e. more than two orders of magnitude higher.  The contamination of $^{214}$Bi is less critical since the limit on the material purity is larger by a factor of $\sim3$ with respect to $^{208}$Tl: as can be seen in Fig.~\ref{fig:extSS} an activity smaller than 80~$\mu$Bq/kg is required for a liquid scintillator thickness of 1.5~m.\\
Given the importance of the purity of the vessel material in terms of $^{208}$Tl contamination, the use of stainless Steel does not seem a viable option and the best option would be to use Copper as foreseen for the Xenon vessel. Considering that an activity of 10~$\mu$Bq/kg can be achieved in Copper, with the proposed liquid scintillator thickness of 1.5~m we could afford to have a thicker vessel: even with a 5~cm thick sphere the background contribution would be below the limit of 0.1 events per year.

\subsection{External gammas background}
\label{sec:ExtGamma}

\begin{figure}[t]
\begin{center}
\includegraphics[width=0.7\textwidth]{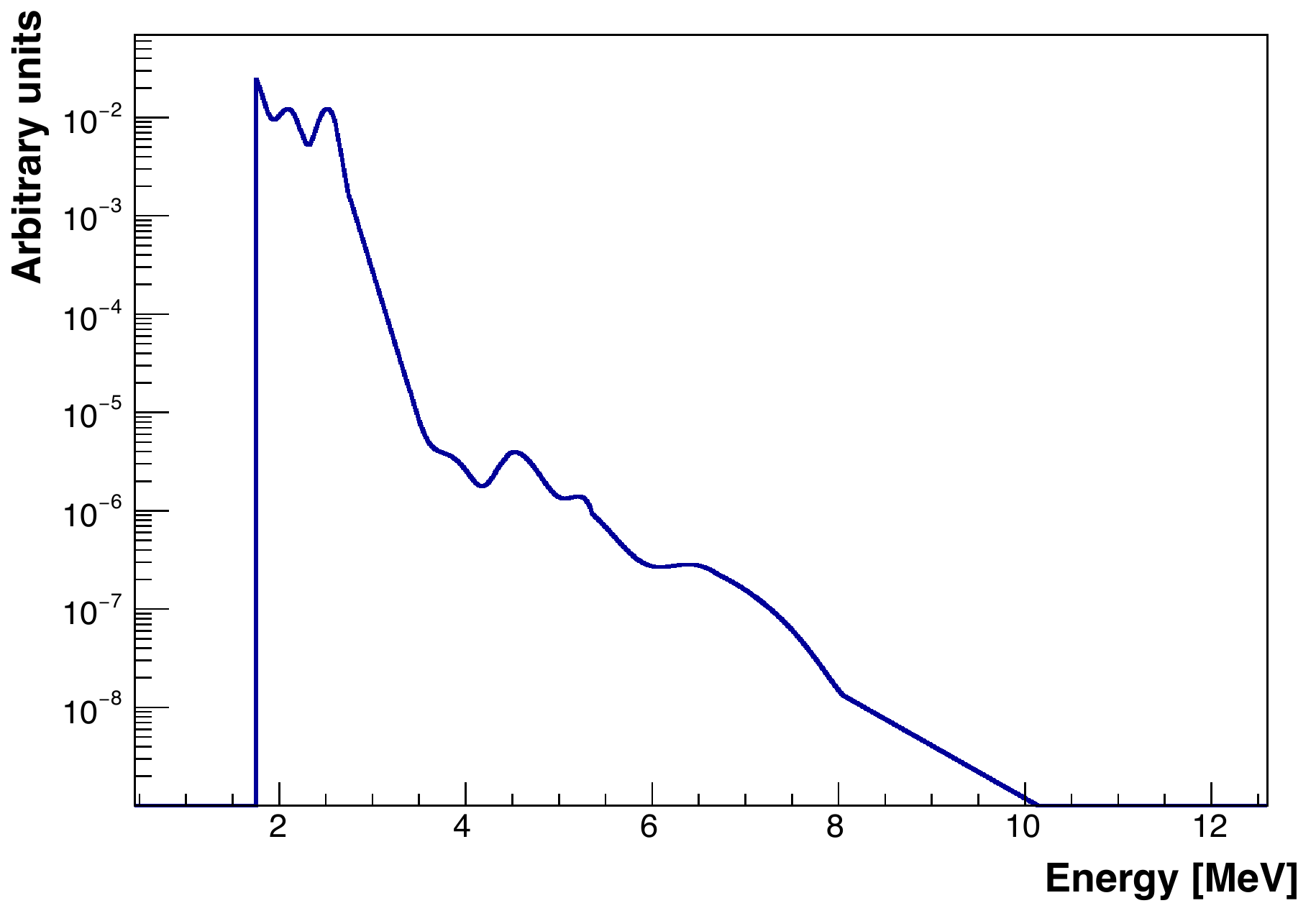}
\caption{{\it External gamma background spectrum in arbitrary units taken from Ref.~\cite{Ohsumi:2002ah}.}}
\label{fig:BGspectrum}
\end{center}
\end{figure}

External gammas present in the laboratory represent another important source of background. A dedicated simulation was performed to evaluate the needed shielding in order to have a background contribution below 0.1 events per year. The simulation is based on the external gamma energy spectrum and rate measured at the LSM laboratory as explained in Ref.~\cite{Ohsumi:2002ah}.\\
It has to be noted that the external background, as highlighted in Ref.~\cite{Ohsumi:2002ah}, depends on the environment itself and a difference up to a factor of 6 was found with respect to measurements carried out in other laboratories. In addition the gamma measurements, in particular in the energy range between 4 and 6~MeV, are detector dependent: despite the effort done by the authors to obtain an absolute measurement subtracting the detector component, the developed simulation could be affected by possible uncertainties in the used rates and spectra. Therefore, the results presented in this section should be taken as an order of magnitude of the background estimate, considering that an error up to a factor of 10 is not impossible.\\
The external gamma spectrum used can be seen in Fig.~\ref{fig:BGspectrum}~\cite{Ohsumi:2002ah}. The background evaluation was carried out for three energy ranges separately: 1.7--4~MeV, 4--6~MeV and 6--10~MeV.\\

\subsubsection{Energy range 1.7--4~MeV}
\label{sec:GLE}
The low energy range is mainly made of gammas produced by natural radioactivity in the materials surrounding the detector.\\
The measured $^{208}$Tl peak rate with no shielding is $4 \times 10^{-2}$ events cm$^{-2}$ s$^{-1}$: considering that it accounts for about 33.5\% of the gamma spectrum in the energy range considered, the total flux used in the MC is 0.12 events cm$^{-2}$ s$^{-1}$.\\
A reduction of $\sim750$ on the Thallium peak was observed with a shielding made of 10~cm of Lead plus 5~cm of Copper, resulting in a total rate of $1.6 \times 10^{-4}$ events cm$^{-2}$ s$^{-1}$. As can be seen in Tab.~\ref{tab:ExtBG}, where the number of events passing the selection cuts is reported, the considered shielding is not enough to reduce the external gamma background to a negligible level. Note that such a shielding was not directly simulated in the MC but it was simply accounted for in the definition of the total expected rate.\\
Two more configurations were studied, where an additional shielding of 5 and 10~cm Lead respectively was added in the MC as an additional spherical layer. The outcome of this study is that to reduce the external gamma background to a level of 0.1 events per year about 25~cm of shielding (total thickness of Lead plus Copper) are needed.

\begin{table}[h]
\begin{center}
\begin{tabular}{|c|c|}
\hline
Shielding & Events per year\\
\hline
No shielding & $20250 \pm 4700$\\
(5~cm Cu + 10~cm Pb)  & $27 \pm 6$\\
(5~cm Cu + 10~cm Pb) + 5~cm Pb & $1.5 \pm 0.4$\\
(5~cm Cu + 10~cm Pb) + 10~cm Pb & $0.12 \pm 0.03$\\
\hline
\end{tabular}
\end{center}
\caption{{\it Number of external gamma background events per year, in the energy region 1.7--4~MeV, for different shieldings. The shielding quoted in parentheses is accounted for in the total rate determination, whereas the one quoted outside the parentheses is directly implemented in the MC as an additional spherical layer. }}
\label{tab:ExtBG}
\end{table}%

\subsubsection{Energy range 4--6~MeV}
\label{sec:GME}
In the presently considered energy range, a large contribution of gammas due to the detector itself is present. A detailed analysis to remove such a contribution was carried out in Ref.~\cite{Ohsumi:2002ah} and the expected rate due to real external gammas is $3.8 \times 10^{-6}$ events cm$^{-2}$ s$^{-1}$.\\
Since the results presented in Sec.~\ref{sec:GLE} on the rejection of low energy gammas showed that a shielding of $\sim 25$~cm Lead is needed, such a setup was used as baseline.\\
As mentioned before only 10~cm Lead are actually simulated in the MC since the remaining 10~cm Lead and 5~cm Copper should be considered in the total rate determination. However, the rate assumed for external gammas in the current energy range does not include such a reduction factor (value not present in Ref.~\cite{Ohsumi:2002ah}). The obtained background rate is therefore conservative since the considered setup would have an additional shielding (about 15~cm more) needed to reject low energy gammas.\\
Out of the $10^{9}$ events simulated, corresponding to a statistics of about 18 years, no event passed the selection cuts. Assuming a maximum allowed number of events of 2.3 with a C.L. of 90\% using Poissonian statistics, a conservative upper limit of 0.13 events per year was obtained.

\subsubsection{Energy range 6--10~MeV}
\label{sec:GHE}
Gammas at energy higher than 6~MeV are mostly due to neutron captures in the surrounding materials.
The analysis is very similar to the one of the 4--6~MeV region, and the expected rate is $3.2 \times 10^{-6}$ events cm$^{-2}$ s$^{-1}$.\\
Looking at the $10^{9}$ events simulated with 10~cm Pb shielding (conservative as explained in Sec.~\ref{sec:GME}), no event remained after the selection cuts. Considering that a statistics of about 21 years was simulated, the Poissonian upper limit of 2.3 events at 90\% C.L. results into a background upper limit of 0.11 events per year.

\subsection{Neutron capture and high energy spallation neutron backgrounds}
\label{sec:NeutronCapture}
\begin{figure}[t]
\begin{center}
\includegraphics[width=0.7\textwidth]{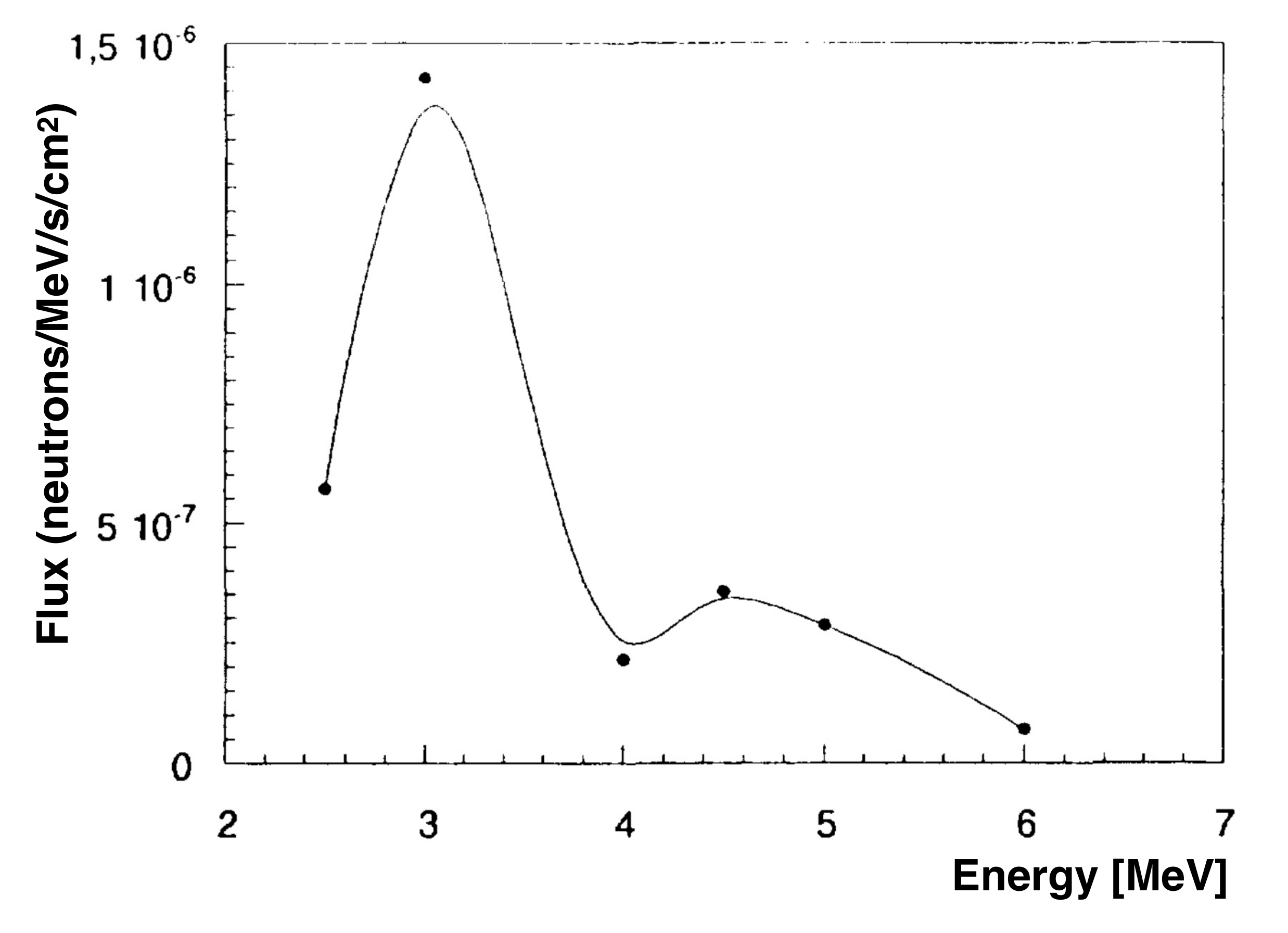}
\caption{{\it External neutron flux measured at LSM taken from Ref.~\cite{NLSM}.}}
\label{fig:NLSM}
\end{center}
\end{figure}

\begin{table}[t]
\begin{center}
\begin{tabular}{|c|c|c|c|c|}
\hline
 & Fraction & Fraction & Mean number   & Mean total  \\
Isotope &  of isotope  &  of  &  of  &  gamma \\
 & abundance &  captures &   gammas &   energy\\
\hline
$^{128}$Xe & 1.91\% & 0.3\% & 5.5  & 6.9 MeV\\
$^{129}$Xe & 26.4\% & 21.5\% & 9.5  & 9.2 MeV\\
$^{130}$Xe & 4.07\% & 0.6\% & 6.4 & 6.6 MeV\\
$^{131}$Xe &  21.23\%& 77.0\% & 4.1 & 7.1 MeV\\
$^{132}$Xe &  26.91\%& 0.4\% & 4.8  & 6.4 MeV\\
$^{134}$Xe &  10.44\%& 0.1\% & 6 &  6.4 MeV\\
$^{136}$Xe &  9.04\% &  0.1\% & 2.1 &  4.0 MeV\\
\hline
\end{tabular}
\end{center}
\caption{{\it Results obtained from the GEANT4 simulation on the fraction of neutron absorption on each Xenon isotope normalized to the total number of absorptions. The mean number of gammas emitted and the mean total energy are also given.}}
\label{tab:ncapture}
\end{table}%

Neutrons produced by spallation of cosmic muons in the sourroundings of the detector, or issued by the radioactive decays and ($\alpha$,n) reactions in the laboratory rock, could results in several types of background.
\begin{itemize}
\item  Neutrons could be captured on different materials present in the laboratory such as Pb, Cu or Fe, and produce high energy gammas in the region of 6 to 10~MeV. Such a contribution has already been accounted for in the external gammas study reported in Sec.~\ref{sec:GHE}.
\item Neutrons could be produced inside the detector by cosmic muons. Such a background can be considered negligible since muons entering the detector are very well tagged, and the low rate expected in underground laboratories (e.g. at LSM the cosmic muon rate is of 4 m$^{-2}$ day$^{-1}$) is small enough to apply a veto after each muon. Thus, neutrons produced by the muon itself can be rejected, as well as the ones issued by decays of cosmogenic isotopes possibly produced, without resulting in any sizeable detector dead time.
\item Externally produced neutrons could enter the detector and be captured in the active volume on the different Xenon isotopes, with the emission of gammas having a total energy of few MeV. Such a background is specifically considered in this section.
\item Spallation neutrons can be produced at energies as high as few GeV and the nuclear recoils could result into signals in the ROI difficult to reject. High energy neutrons are also dealt with in this section.
\end{itemize}

To estimate the background, as far as neutron captures are concerned, the active volume of the GEANT4 simulation was defined considering the natural abundance of the different Xenon isotopes without assuming any enrichment in $^{136}$Xe. The results are therefore conservative since the neutron capture cross section of $^{131}$Xe, one of the most abundant isotope, is larger by more than two order of magnitudes with respect to the one of $^{136}$Xe, as can be inferred from Tab.~\ref{tab:ncapture}.\\
The external neutron flux of $4 \times 10^{-6}$ events cm$^{-2}$ s$^{-1}$ and their energy spectrum (as shown in Fig.~\ref{fig:NLSM}) are taken from measurements performed at LSM~\cite{NLSM}. A statistics of $6 \times 10^8$ events was generated outside the Pb shielding, equivalent to a data taking of $\sim 10$~years.
 Since no specific shielding for neutrons such as Polyethylene was considered, and only 10~cm of Pb were used in the MC although at least 25~cm are needed to reject low energy external gammas (see Sec.~\ref{sec:GLE}), the obtained results can be considered as an upper limit and highly conservative.\\
The neutron capture cross sections as well as the gamma de-excitation cascades on the different Xenon isotopes are already included in the GEANT4 toolkit. In Tab.~\ref{tab:ncapture} the fraction of neutrons absorbed on each Xenon isotope with the mean number of gammas emitted and the mean total energy are shown.\\
The analysis on the simulated events resulted in no events selected as signal which at 90\% C.L. give an upper limit of 2.3 events i.e. less than 0.23 events per year.

To study high energy neutrons the energy spectrum from Ref.~\cite{Mei:2005gm} was used in the MC generator, and a flux above 10~MeV of $0.5 \times 10^{-9}$~cm$^{-2}$~s$^{-1}$ was assumed, which is a conservative estimate interpolating between the expectation for Gran Sasso underground laboratory and Sudbury one, at an underground depth of 4800 m.w.e. corresponding to the LSM laboratory.\\
A statistics corresponding to 11 years of exposure was simulated outside the 25~cm Lead shielding: zero events were found in the ROI before applying any additional selection cut. The upper limit at 90\% C.L. corresponds therefore at 0.21 events per year.

\subsection{Radon background}
\label{sec:RnBG}
The background related to $^{222}$Rn has to be studied in detail. Given the property of such a gas to stick on different materials, the contribution to the background due to its presence inside the Xenon gas or in the liquid scintillator has to be evaluated. In addition $^{222}$Rn quickly decays into $^{210}$Pb which has a half-life of about 22~years and which could result into an almost constant source of background over the whole data taking.\\
The first point to assess is the maximal amount of Radon which is acceptable inside the detector to keep the background at a level of less than 0.1 events per year. Assuming a constant probability per unit volume, the Radon decay chain was simulated inside the Xenon active volume and the full analysis chain was applied to the MC events. Note that, as stated in Sec.~\ref{sec:simu}, a time window of 10~ms is used to separate events in the decay chain therefore the Bi-Po decays are seen as a single event in the current detector simulation, and no quenching for $\alpha$'s is considered.\\ 
Considering the fraction of events resulting into background is at the level of $2 \times 10^{-7}$, the maximum amount of Radon allowed in the active Xenon corresponds to an activity of 50~$\mu$Bq/kg, or equivalently to 12 mBq/m$^{3}$.\\
The Radon contamination in the liquid scintillator is less critical in terms of background: in more than 80\% of the events there is an energy deposition in the liquid scintillator which is larger than the threshold of 200~keV used for the background rejection. The MC showed indeed that, at the same level of Radon contamination, the events issued in the liquid scintillator resulting into background are suppressed by more than one order of magnitude with respect to the ones issued in the active volume itself.\\
 
\subsection{Cosmogenic background}
Muons crossing the Xenon volume could produce cosmogenic nuclei inside the active volume. Some of them might result into a source of background since they are unstable and undergo beta decays which might yield electrons with energies in the ROI. In particular several isotopes of Iodine ($^{130}$I, $^{132}$I, $^{134}$I, $^{135}$I) and Xenon ($^{137}$Xe) were considered based on the cosmogenic background measurements carried out by the EXO-200 experiment~\cite{EXO200::2015wtc}.\\
The largest source of background comes from $^{137}$Xe whereas the contribution of Iodine isotopes is smaller by more than two orders of magnitude. After all selection cuts, about 0.18\% of the $^{137}$Xe events, uniformly generated inside the Xenon active volume, result into background. To estimate the rate of events per year, the number of produced $^{137}$Xe cosmogenic nuclei is needed, which depends on the muon rate and on the production yield. A way to compute this is to use the production yield of 440 events per year measured by EXO-200 and rescale it by the Xenon mass i.e. by a factor of $\sim 3.4$. This is very conservative since the muon rate at LSM, where we assume to have our detector, is smaller by more than a factor 100 with respect to the muon rate measured at Waste Isolation Pilot Plant (WIPP) where the EXO-200 measurements were carried out. With the assumed $^{137}$Xe production rate the background contribution is of 0.2 events per year, without taking into account the reduction due to the larger overburden. Considering that the production yield depends on the muon energy and that it might be a factor 4 higher at LSM with respect to WIPP~\cite{Mei:2005gm}, the fact that the muon rate is about 100 times smaller results into a reduction of $\sim 25$ i.e. less than 0.01 events per year.

\section{Event rate in liquid scintillator}
\label{sec:LSrate}
The liquid scintillator veto provides an important tool for background rejection as explained in Sec.~\ref{sec:BGrejection}, however there is a possible drawback: if the rate of events triggering the scintillator is too high, this would result into a signal inefficiency.\\
All the possible sources contributing to the trigger rate in the liquid scintillator were considered as reported in Tab.~\ref{tab:LStrigger}.
In addition to the radioactive decays in the different spherical vessels, and the previously studied external gammas and neutrons, external gammas below 1.7~MeV were studied. They do not represent an issue in the background study due to their energy well below the ROI lower boundary, but they might have an impact on the total trigger rate of the liquid scintillator.
Indeed they represent an issue if the shielding is not thick enough (38~Hz for 10~cm Pb + 5~cm Cu), however when the shielding is at the level of $\sim 25$~cm of Lead as needed for the background reduction (see Sec.~\ref{sec:GLE}) the rate drops to 0.02~Hz.\\
The largest contribution to the trigger rate comes from the radioactive decays in the vessel containing the liquid scintillator. The contribution of the full $^{238}$U and $^{232}$Th decay chains as well as the one of the $^{60}$Co decay was evaluated. Their contributions are similar, yielding a total trigger rate in the liquid scintillator of about 0.9~Hz.\\
The Radon contribution deserves a specific discussion. A limit on the maximal allowed activity of $\sim 50$~$\mu$Bq/kg has been set in Sec.~\ref{sec:RnBG} based on the desired maximal background of 0.1 events per year. This was driven by the contamination in the Xenon: the background issued by the Radon contamination in the liquid scintillator is indeed strongly suppressed by the veto provided by the liquid scintillator itself. The limit on the maximal accepted Radon activity based on the liquid scintillator trigger rate is however much more constraining. To keep the liquid scintillator trigger rate contribution below 1~Hz (i.e. the contribution due to the vessel) the maximal activity allowed is about 5~$\mu$Bq/kg, which is one order of magnitude smaller than the allowed contamination inside the Xenon volume based on the background rate. Despite the high purity required, an activity of 5~$\mu$Bq/kg is a factor of $\sim 100$ higher that what has been already achieved in other detectors (e.g. the inner balloon of the KamLAND experiment~\cite{Obara:2017ndb}), therefore it does not represent a show stopper for the considered detector design.\\
An additional contribution to the trigger rate of the liquid scintillator comes from the $^{210}$Pb surface contamination of the vessel if exposed to Radon before the detector assembly. Setting again as a limit a rate of 1~Hz and considering the vessel dimensions, a total deposition of about 1600 Radon daughters per cm$^2$ or an equivalent activity of 2~$\mu$Bq/cm$^2$ can be afforded. This limit is one order of magnitude larger than what has already been achieved in CUORE~\cite{Arnaboldi:2002du} and in the setup at LSM~\cite{Fard:2015pla} ($^{210}$Pb contamination of the order of 0.15 $\mu$Bq/cm$^2$).\\ 
The total trigger rate amounts to about 2.9~Hz: this is acceptable since assuming events 1~ms long, it would result into a signal inefficiency at the few per-mil level.\\

\begin{table}[t]
\small
\begin{center}
\begin{tabular}{|c|c|c|c|}
\hline
 &   & Fraction of events  & Rate \\
Source & Event rate &  with E $>$ 200~keV  & in  liquid \\
 &   &  in  liquid scintillator & scintillator\\
 \hline
 Cu Xe vessel   & 0.018Hz (U+Th+Co)  &  & $2.3\times 10^{-3}$~Hz (U+Th+Co)  \\
 (0.5~cm thick,   &  0.01~Hz (U)  & 7\%(U)  & $7\times 10^{-4}$~Hz (U)\\
 10~$\mu$Bq/kg)  & $7.2\times 10^{-3}$~Hz (Th)   & 12\%(Th) & $8.6\times 10^{-4}$~Hz (Th)\\
 &  $8\times 10^{-4}$~Hz (Co)  & 94\%(Co) & $7.5\times 10^{-4}$~Hz (Co) \\
\hline
 Stainless Steel/Cu    & 12.8Hz (U+Th+Co)  &  & 0.9~Hz (U+Th+Co)  \\
  LS vessel &  7.1~Hz (U)  & 4\%(U)  & 0.28~Hz (U)\\
 (2~cm thick,  & 5.1~Hz (Th)   & 6\%(Th) & 0.31~Hz (Th)\\
 80~$\mu$Bq/kg)  &  0.55~Hz (Co)  & 59\%(Co) & 0.32~Hz (Co) \\
\hline
 External gammas &  180~Hz &    &  \\
 $[0.6-1.7]$~MeV  &  (after 10~cm Pb  &  0.013\%  &  0.02~Hz\\
 (20~cm Pb + 5~cm Cu)  & + 5~cm Cu) &    &  \\
\hline
 External gammas &  73~Hz  &    &  \\
 $[1.7-4]$~MeV  &  (after 10~cm Pb &  0.11\%  &  0.08~Hz\\
 (20~cm Pb + 5~cm Cu)  & + 5~cm Cu) &    &  \\
\hline
 External gammas &   &    &  \\
 $[4-6]$~MeV  & 1.7~Hz  &  0.2\%  &  $3\times 10^{-3}$~Hz\\
 (10~cm Pb)  &   &    &  \\
\hline
 External gammas &   &    &  \\
 $[6-10]$~MeV  & 1.5~Hz  &  0.16\%  &  $2.4\times 10^{-3}$~Hz\\
  (10~cm Pb)  &   &    &  \\
\hline
$^{222}$Rn &   &    &  \\
in Xenon  & 0.0175~Hz  &  13\%  &  $2.3\times 10^{-3}$~Hz\\
 (assuming 50~$\mu$Bq/kg)  &   &    &  \\
 \hline
$^{222}$Rn &   &    &  \\
in LS  & 0.82~Hz  &  80\%  &  0.66~Hz\\
 (assuming 5~$\mu$Bq/kg)  &   &    &  \\
 \hline
 $^{210}$Pb on LS &   &    &  \\
 vessel surface & 0.88~Hz  &  100\%  &  0.88~Hz\\
 (assuming 2~$\mu$Bq/cm$^2$)  &   &    &  \\
 \hline
External fast &   &    &  \\
 neutrons & 2~Hz  &  18\%  &  0.36~Hz\\
  (10~cm Pb) &   &    &  \\
  \hline
External high energy &   &    &  \\
($>$ 10 MeV) fast neutrons & $2.9\times 10^{-4}$~Hz  &  31\%  &  $9\times 10^{-5}$~Hz\\
  (25~cm Pb) &   &    &  \\
 \hline
Cosmogenic $^{137}$Xe &  $4\times 10^{-6}$~Hz  &  26\%  &  $1\times 10^{-6}$~Hz\\
 \hline
\hline
 Total  & -  &  -  & $\sim 2.9$~Hz \\
 \hline

\end{tabular}
\end{center}
\caption{{\it Trigger rate of the liquid scintillator due to the different possible sources.}}
\label{tab:LStrigger}
\end{table}%

\section{Pile up evaluation}
In addition to the issue represented by the trigger rate in the liquid scintillator, all the possible sources of background could result into pile up.\\
Events depositing energy in the Xenon active volume are dangerous since an additional energy deposition in the signal time window would affect the energy reconstruction. Of course  background events yielding energy deposition in the active Xenon volume should not trigger the liquid scintillator, otherwise a veto would be applied and they would not contribute to the pile up.\\
The expected pile up rates of the different considered sources was computed: as can be seen in Tab.~\ref{tab:pileup} pile up is not an issue with the presently assumed background rates.
The dominant contribution comes from the Radon contamination in the Xenon active volume, however with the assumed Radon contamination in the Xenon gas at the level of 50~$\mu$Bq/kg, the pile up rate is totally negligible being of $\sim 1.5 \times 10^{-2}$~Hz.

\begin{table}[htbp]
\small
\begin{center}
\begin{tabular}{|c|c|c|c|}
\hline
 &   & Fraction of events  & Rate \\
Source & Event rate &  with E$>$0 in Xe   & in Xenon \\
 &   &   and E $<$ 200~keV  &   active \\
 &   &  in  liquid scintillator & volume\\
 \hline
 Cu Xe Sphere   & 0.018Hz (U+Th+Co)  &  & $5.5\times 10^{-4}$~Hz (U+Th+Co)  \\
 (0.5~cm thick,   &  0.01~Hz (U)  & $3\times 10^{-2}$ (U)  & $3\times 10^{-4}$~Hz (U)\\
 10~$\mu$Bq/kg)  & $7.2\times 10^{-3}$~Hz (Th)   & $4.3\times 10^{-2}$ (Th) & $3.1\times 10^{-4}$~Hz (Th)\\
 &  $8\times 10^{-4}$~Hz (Co)  & $5\times 10^{-2}$ (Co) & $4\times 10^{-5}$~Hz (Co) \\
\hline
 Stainless steel/Cu    & 12.8Hz (U+Th+Co)  &  & $<2.9\times 10^{-4}$~Hz (U+Th+Co)  \\
  LS Sphere &  7.1~Hz (U)  & $<2.3\times 10^{-5}$ (U)  & $<1.6\times 10^{-4}$~Hz (U)\\
 (2~cm thick,  & 5.1~Hz (Th)   & $<2.3\times 10^{-5}$ (Th) &$<1.2\times 10^{-4}$~Hz (Th)\\
 80~$\mu$Bq/kg)  &  0.55~Hz (Co)  & $<2.3\times 10^{-5}$ (Co) & $<1.3\times 10^{-5}$~Hz (Co) \\
\hline
 External gammas & 180~Hz  &    &  \\
 $[0.6-1.7]$~MeV  & (after 10~cm Pb    &  $<2.3\times 10^{-6}$  &  $<4.1\times 10^{-4}$~Hz\\
 (20~cm Pb + 5~cm Cu)  & + 5~cm Cu)&    &  \\
\hline
 External gammas &  73~Hz  &    &  \\
 $[1.7-4]$~MeV  &  (after 10~cm Pb &  $<2.3\times 10^{-6}$  &  $<1.4\times 10^{-4}$~Hz\\
 (20~cm Pb + 5~cm Cu) & + 5~cm Cu)  &    &  \\
\hline
 External gammas &   &    &  \\
 $[4-6]$~MeV  & 1.7~Hz  &  $<2.3\times 10^{-6}$  &  $<3.9\times 10^{-6}$~Hz\\
 (10~cm Pb)  &   &    &  \\
\hline
 External gammas &   &    &  \\
 $[6-10]$~MeV  & 1.5~Hz  &  $<2.3\times 10^{-6}$  &  $<3.5\times 10^{-6}$~Hz\\
 (10~cm Pb)  &   &    &  \\
\hline
$^{222}$Rn &   &    &  \\
in Xenon  & 0.0175~Hz  &  0.86  &  $1.5\times 10^{-2}$~Hz\\
 (assuming 50~$\mu$Bq/kg)  &   &    &  \\
 \hline
$^{222}$Rn &   &    &  \\
in LS  & 0.82~Hz  &   $<2.3\times 10^{-5}$  &  $<1.9\times 10^{-5}$~Hz\\
 (assuming 5~$\mu$Bq/kg)  &   &    &  \\
  \hline
  
 $^{210}$Pb on LS &   &    &  \\
 vessel surface & 0.88~Hz  &  $<2.3\times 10^{-5}$   &  $<2.0\times 10^{-5}$~Hz\\
 (assuming 2~$\mu$Bq/cm$^2$)  &   &    &  \\
\hline
 External fast &   &    &  \\
neutrons & 2~Hz  & $<2.3\times 10^{-5}$  &  $<4.6\times 10^{-5}$~Hz\\
(10~cm Pb) &   &    &  \\
 \hline
External high energy &   &    &  \\
($>$ 10 MeV) fast neutrons & $2.9\times 10^{-4}$~Hz  &  $<2.3\times 10^{-4}$  &  $<6.7\times 10^{-8}$~Hz\\
  (25~cm Pb) &   &    &  \\
 \hline 
 Cosmogenic $^{137}$Xe &  $4\times 10^{-6}$~Hz  &  0.74  &  $3\times 10^{-6}$~Hz\\
 \hline
 \hline
 Total  & -  &  -  & $\sim1.5\times 10^{-2}$~Hz \\
 \hline
\end{tabular}
\end{center}
\caption{{\it Trigger rate in the Xenon active volume due to the different possible background sources.}}
\label{tab:pileup}
\end{table}%

\section{Light readout}
\begin{figure}[t]
\begin{center}
\includegraphics[width=0.7\textwidth]{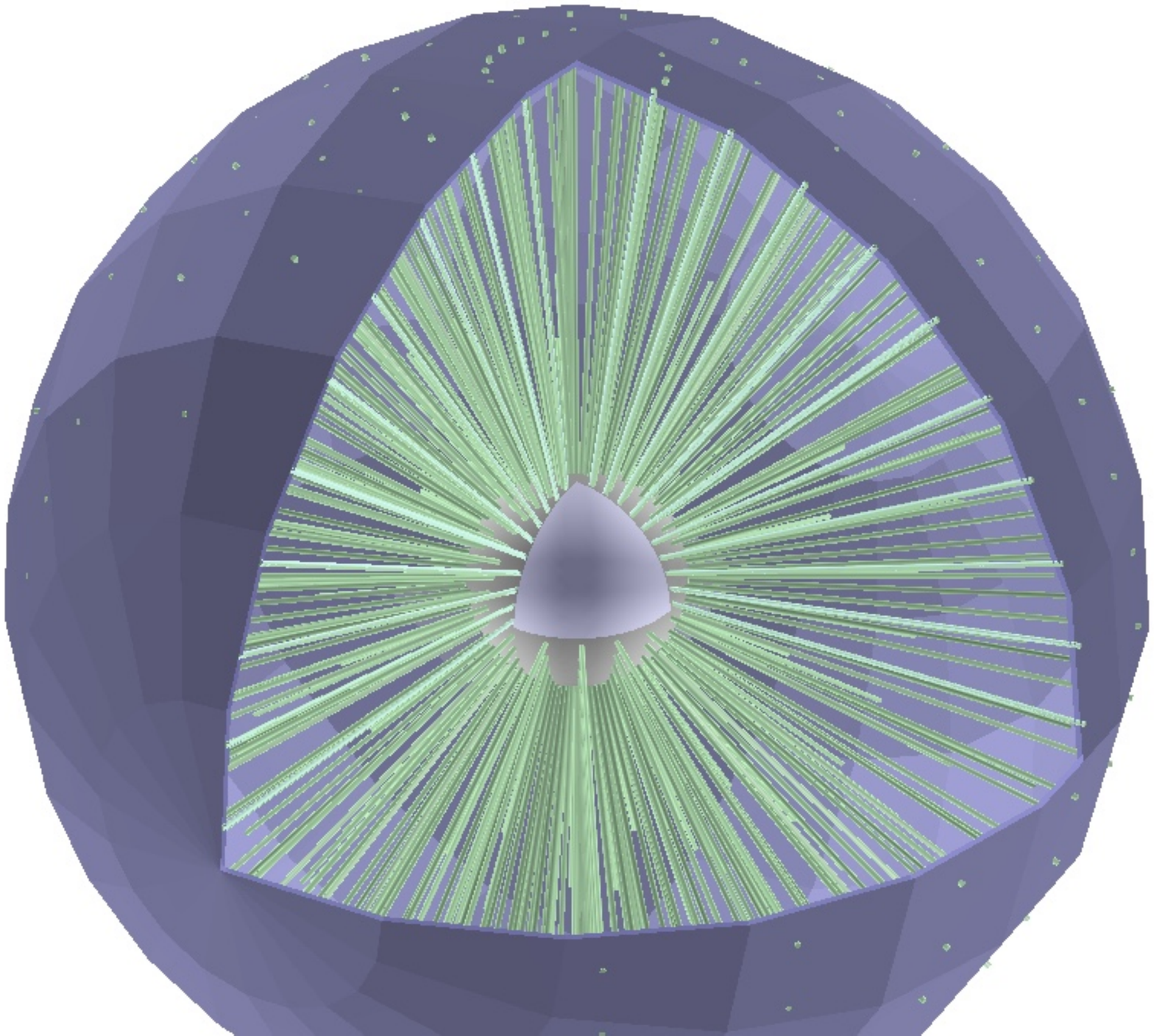}
\caption{{\it Cartoon showing the WLS fibers (400 in this example) connecting the Xenon vessel (inner light grey sphere) to the liquid scintillator vessel (outer violet sphere).}}
\label{fig:WLS}
\end{center}
\end{figure}

\begin{figure}[tbp]
\begin{center}
\includegraphics[width=0.7\textwidth]{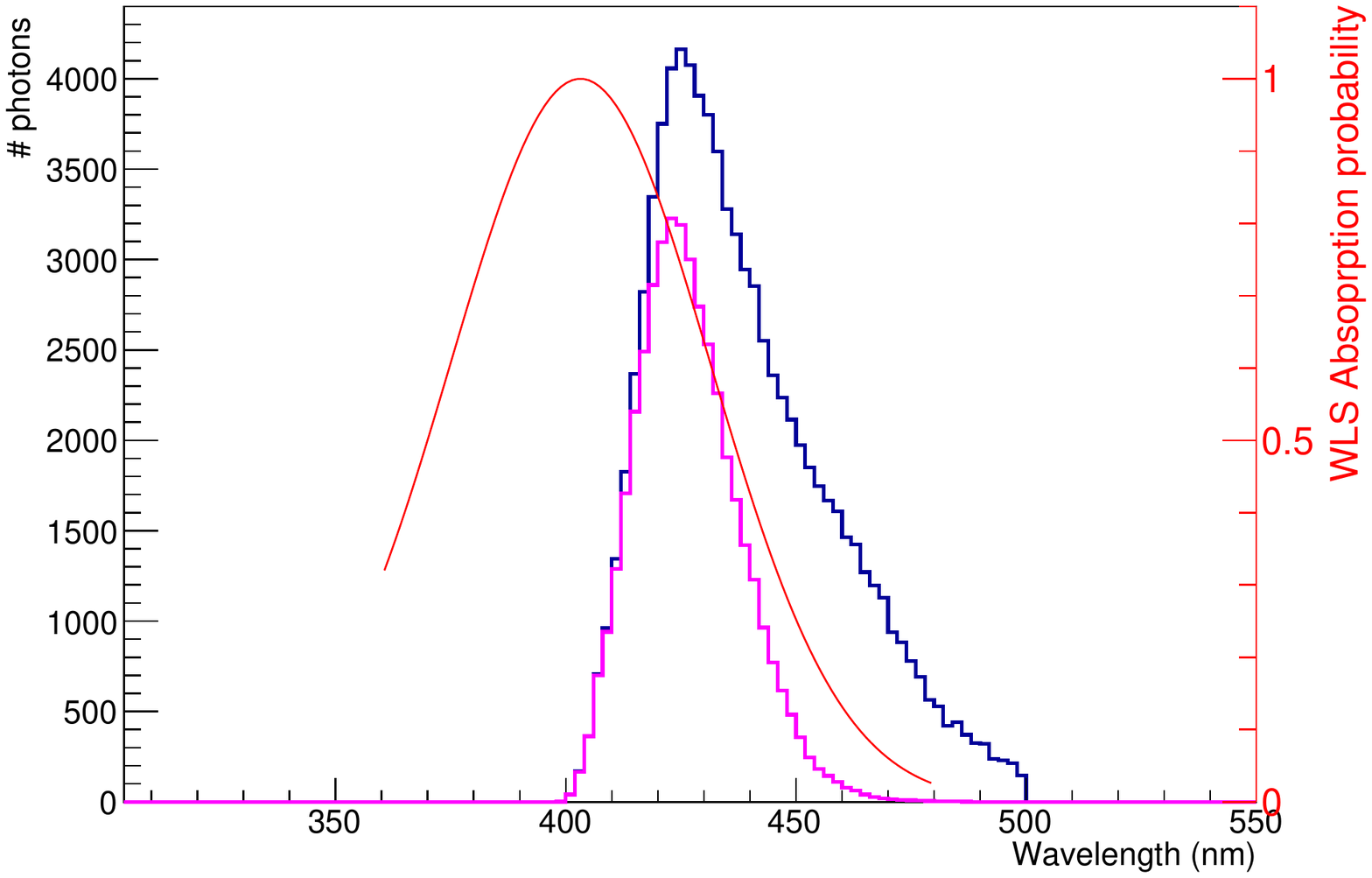}
\caption{{\it Spectra of photons reaching the WLS fibers (blue histogram) and spectra of photons absorbed (purple histogram) after the convolution with the absorption probability (red curve with respect to the right vertical axis as taken from Ref.~\cite{SG}).}}
\label{fig:WLS2}
\end{center}
\end{figure}

In the background rejection it has been assumed that it is possible to observe energy deposits as small as 200~keV in the liquid scintillator. The feasibility of such a measurement is addressed in this section.\\
The liquid scintillator has to be instrumented with photomultipliers tubes (PMTs) in order to detect the scintillation light emitted by the particles depositing energy in the liquid scintillator.  However, the presence of PMTs in the liquid scintillator results into an additional event rate increasing the possible accidental coincidence with a signal, and therefore the inefficiency. Even low radioactivity PMTs such as the ones of SuperNEMO~\cite{Ohsumi:2008zz}, have an activity at the level of 1~Bq per PMT which would increase by far the estimated trigger rate in the liquid scintillator.\\
A different approach was therefore considered i.e. the use of wavelength shifting (WLS) fibers which can collect the scintillation light inside the liquid scintillator and transport it outside the metal vessel where they could be grouped in bundles and read by photodetectors.
To assess the feasibility of such a readout, an optical simulation was included in the MC setup simulation where the fibers are distributed radially from the Xenon spherical vessel to the liquid scintillator one in an isotropic way as shown in Fig.~\ref{fig:WLS}.\\
The following assumptions were made in the simulation:
\begin{itemize}
\item Energy deposits of 200~keV were simulated in the liquid scintillator volume assuming a flat volume distribution (i.e. same probability per unit volume).
\item A light yield of 10000 photons per MeV was assumed~\cite{LS}.
\item An absorption length of 20~m was assumed~\cite{Yang:2017ybj}.
\item A reflection probability of 50\% of optical photons on the metal vessel surfaces was assumed.
\item BCF-92 WLS fibers~\cite{SG} with a diameter of 2~mm were considered.
\item The emission spectrum of the scintillator (modelled as a standard plastic scintillator since at this point the real chemistry of the LAB based scintillator is not defined) as well as the absorption one of the WLS fibers was taken into account as taken from Ref.~\cite{SG}.
\item All the optical photons reaching the fibers were counted as absorbed (the convolution of the absorbed spectrum photons with the real WLS absorption spectrum was done at the analysis level).
\item An efficiency of 10\% is assumed for the emitted photons to undergo total reflection inside the WLS fiber.
\item The attenuation length inside the WLS fiber was not simulated since the effect is expected to be very small (attenuation length of about 10~m with respect to a maximal distance travelled by photons in the current setup of 1.5~m).
\item A quantum efficiency of 25\% was assumed in the conversion of the light to photo-electrons (p.e.).
\end{itemize}
The density of the WLS fibers was changed keeping an isotropic distribution and increasing the total number in a scan going from a total of 100 fibers up to 2500.
Convoluting the spectrum of the emitted optical photons to the absorption spectrum of the BCF-92 WLS fibers, the net result is a reduction of a factor of $\sim 2$ on the number of detected photons as can be seen in Fig.~\ref{fig:WLS2}.\\
The resulting number of photo-electrons for the different configurations is summarized in Tab.~\ref{tab:WLS}. If 4 p.e. is considered a reasonable threshold to detect the signal, a configuration with about 2000 fibers is sufficient. The minimal number of p.e. in this configuration is 2.5 which is still large enough to assure an almost negligible veto trigger inefficiency. In addition some room of improvement is still possible since the scintillator emission spectrum could be better matched to the fiber absorption one.\\
Note that the number of fibers does not correspond to the number of readout channels since fibers would be grouped together in bundles to be read by the same PMT. Indeed the number of PMTs has to be limited in order to avoid a high coincidence rate due to dark current: assuming a dark counting rate of the order of 200~Hz and a time window of 20~ns we should not exceed $\sim50$ PMTs to have a dark counting coincidence of about 1~Hz. Another possible solution would be the use of multi anode PMTs which have low dark counting rate at the level of 3~Hz per channel~\cite{Adam:2007ex}.\\

\begin{table}[htbp]
\begin{center}
\begin{tabular}{|c|c|}
\hline
Number of fibers & mean  number of p.e.\\
\hline
100 & 0.3\\
400 & 1.0\\
900 & 2.2\\
1600 & 3.5\\
2025 & 4.3\\
2500 & 4.6\\
\hline
\end{tabular}
\end{center}
\caption{{\it Mean number of p.e. observed for a 200~keV energy deposit, for different WLS fiber configurations.}}
\label{tab:WLS}
\end{table}%

A different light readout is also under investigation using transparent windows in the LS vessel  with light guides on the outer part connected to PMTs.\\
The same MC simulation was used and round windows with a radius of 10~cm were considered. Under the same liquid scintillator properties as detailed before, considering a quantum efficiency of 25\% of the PMT and no light loss in the light guide between the transparent window and the PMT, the number of p.e. which could be observed was computed as a function of the number of windows.\\
With 16 transparent windows we achieved a mean number of p.e. of 4.6 and a minimum number of p.e. of 1.8 which is comparable to the performance obtained using WLS fibers. The advantage of such a light readout is the reduced number of channels and the simpler detector layout.

\section{Summary on the detector constraints}
\label{sec:constraint}
In the previous sections several constraints on the detector arose either from the requirement of a very low background, or from the allowed limit on the liquid scintillator trigger rate.\\
None of them turned into a show stopper for the proposed experimental setup, nonetheless most of them demand a particular care in the choice of materials or on their handling.
The most important points arisen in the previously presented studies are summarized in Tab.~\ref{tab:constraint}.

\begin{table}[htbp]
\begin{center}
\begin{tabular}{|c|c|}
\hline
Constraint  & Reason \\
\hline
Copper Xenon vessel activity & Background from the Xenon vessel\\
below 10~$\mu$Bq/kg & at the level of 1 event per year in 50 kg\\
\hline
Liquid scintillator & \\
thickness of at least 150~cm & Reduce the background from the liquid  \\
 & scintillator vessel below the level of\\

Copper/stainless steel liquid &  0.1 events per year in 50 kg \\
scintillator vessel activity  &  \\
 below 25~$\mu$Bq/kg & \\
 \hline
 & Reduce the background from external gammas  \\
 Lead shielding of &  below the level of 0.1 events per year in 50 kg  \\
 at least 25~cm &\\
&  Reduce the trigger rate in LS due to low energy  \\
 &  external gamma below the level of 1~Hz \\
 \hline
 Radon activity in Xenon volume & Reduce the background from Radon \\
 below 50~$\mu$Bq/kg & below the level of 0.1 events per year in 50 kg \\
  \hline
Radon activity in LS volume & Reduce the trigger rate in LS  \\
 below 5~$\mu$Bq/kg &due to Radon below the level of 1~Hz \\
\hline
Lead activity on LS vessel & Reduce the trigger rate in LS  \\
 surface below 2~$\mu$Bq/cm$^2$ &due to $^{210}$Pb below the level of 1~Hz \\
\hline
\end{tabular}
\end{center}
\caption{{\it Major constraints on the detector setup arisen from the presented studies.}}
\label{tab:constraint}
\end{table}%

\section{Outlook}
In the present paper we have decided to be conservative under many aspects and assess the proof of principle of a detector which could be built with the presently available technology. Nonetheless many features could be improved resulting into a further reduction of the background. Here below we report a non-exhaustive list of the different items under investigation which could be improved possibly resulting into a lower background or a more compact detector.\\
\begin{itemize}
\item The activity assumed for Copper is at the level of 10~$\mu$Bq/kg whereas Copper with an activity ten times lower is available on the market. Such a reduction would result not only into a smaller background but also into the possibility to use a thinner volume of liquid scintillator for the veto reducing the size of the detector and therefore of the shielding.\\
\item The energy resolution is a critical point: we assumed 1\% FWHM at the Xenon $Q_{\beta\beta}$ (i.e. 2.458~MeV) with a gas pressure of 40 bar. This has to be demonstrated, however previous measurements performed on Xenon TPC using $^{137}$Cs source showed that a resolution of 0.4\% FWHM can be achieved at 662~keV for Xenon densities below 0.6 g/cm$^3$~\cite{Bol:1997}. 
Test to measure the energy resolution in a spherical TPC filled with pure Xenon are ongoing at LSM whereas the drift and the amplification of the charge has already been demonstrated up to a few bars.\\
\item The use of multi-ball central anode~\cite{Giganon:2017isb} is under investigation. The advantage would be a  reduction of the high voltage, which could be a limiting factor when the sphere size gets too large, and the possibility to perform a coarse tracking which would be an additional handle in the background reduction. This is particularly true for $^{214}$Bi induced background which is the result of multi-gamma events: a cut on a 3D energy deposition, more efficient than the $\Delta$R cut applied in the performed analysis, seems to reduce the $^{214}$Bi background by about 30\%. Combined with recent developments in the waveform processing of the sphere TPC signal, a reconstruction of each individual ionization track might also be obtained.\\
\item The available optical simulation showed that it is possible to read the light of the liquid scintillator detecting energy deposits of 200~keV. Further tuning is possible: a more detailed simulation, taking into account for example the diffusion of photons by Teflon which could be used to wrap the outer part of the Xenon vessel, will be developed to optimize the detector design and maximise the number of detected photo-electrons.\\
\item If the scintillation light of the Xenon gas is detected, we would have a precise time stamp of the event allowing to make a coincidence with the liquid scintillator signal on a much smaller timescale and therefore reducing the accidental rate. In addition a precise time stamp of the event would result into a better radial reconstruction of the position. Different way of reading the Xenon light are under study including the possibility of having additional wavelength shifting fibers inside the Xenon vessel, or depositing some cathode on the inner source of the vessel to read the emitted electrons with the detector itself after their drift towards the central anode.
\end{itemize}

\section{Conclusions}
A possible setup for a spherical high pressure Xenon gas TPC for the search of $\beta\beta0\nu$ decay has been studied.\\
The presented setup consists of an inner Xenon volume and an outer volume made of liquid scintillator which is used as veto to reduce the background. Under the assumption of an energy resolution of 1\% FWHM at the $Q_{\beta\beta}$ of 2.458~MeV, an optimized ROI of $Q_{\beta\beta} \pm 0.6\%$, the possibility of a radial energy deposition reconstruction and a threshold of 200~keV for the liquid scintillator, a background of about 2 events per year was obtained with a signal efficiency of $\sim 64\%$ and a limit of $T_{1/2}^{0\nu} > 2.5 \times 10^{25}$ years.\\
The obtained results assume some constraints on the allowed activities of the different materials as summarized in Sec.~\ref{sec:constraint}, although none of them represents a critical issue.\\
Different possible backgrounds were studied showing that all of them can be reduced to the negligible level of 0.1 events per year in the detector mass of 50~kg.\\
The trigger rate of the liquid scintillator was evaluated resulting in a total signal inefficiency below 0.4\%.\\
A simulation of the light readout, studying different possible options, was also performed showing that it is indeed possible to observe, with almost negligible inefficiency, energy depositions inside the liquid scintillator as small as 200~keV.\\
The presented results show that the proposed detector could give competitive results with present experiment for what concerns internal background. This could be the first step of a more challenging roadmap which would include the use of different gases to measure the $\beta\beta0\nu$ signal from different isotopes with the same detector (i.e. the same background sources), and possibly at a higher $Q_{\beta\beta}$ therefore with a smaller background. \\


\begin{thebibliography}{99}

\bibitem{Gerbier:2014jwa} 
  G.~Gerbier {\it et al.},
  arXiv:1401.7902 [astro-ph.IM].

\bibitem{Arnaud:2017bjh} 
  Q.~Arnaud {\it et al.} [NEWS-G Collaboration],
  arXiv:1706.04934 [astro-ph.IM].
  
\bibitem{gdr} 
 I.~Giomataris,
Talk at GDR neutrino meeting 2015 at Saclay.\\  
https://indico.in2p3.fr/event/12176/
  
\bibitem{Auger:2012ar} 
  M.~Auger {\it et al.} [EXO-200 Collaboration],
  Phys.\ Rev.\ Lett.\  {\bf 109}, 032505 (2012).

\bibitem{Lorca:2012dv} 
  D.~Lorca {\it et al.} [NEXT Collaboration],
  Nucl.\ Instrum.\ Meth.\ A {\bf 718}, 387 (2013).

\bibitem{Gando:2012zm} 
  A.~Gando {\it et al.} [KamLAND-Zen Collaboration],
  Phys.\ Rev.\ Lett.\  {\bf 110}, no. 6, 062502 (2013).
  
\bibitem{Savvidis:2016wei} 
  I.~Savvidis, I.~Katsioulas, C.~Eleftheriadis, I.~Giomataris and T.~Papaevangellou,
  arXiv:1606.02146 [physics.ins-det].
  
\bibitem{Agostinelli:2002hh} 
  S.~Agostinelli {\it et al.} [GEANT4 Collaboration],
  Nucl.\ Instrum.\ Meth.\ A {\bf 506}, 250 (2003).
  
\bibitem{Redshaw:2007un} 
  M.~Redshaw, E.~Wingfield, J.~McDaniel and E.~G.~Myers,
  Phys.\ Rev.\ Lett.\  {\bf 98}, 053003 (2007).
  
   \bibitem{BBspectra}
 http://nucleartheory.yale.edu/double-beta-decay-phase-space-factors
 
\bibitem{Kotila:2012zza} 
  J.~Kotila and F.~Iachello,
  Phys.\ Rev.\ C {\bf 85}, 034316 (2012).
  
\bibitem{NuclearData} 
http://nucleardata.nuclear.lu.se/toi/nuclide.asp?iZA=810208

\bibitem{Agashe:2014kda}
  K.~A.~Olive {\it et al.} [Particle Data Group],
  Chin.\ Phys.\ C {\bf 38} 090001 (2014).
  
\bibitem{Maneschg:2008zz} 
  W.~Maneschg, M.~Laubenstein, D.~Budjas, W.~Hampel, G.~Heusser, K.~T.~Knopfle, B.~Schwingenheuer and H.~Simgen,
  Nucl.\ Instrum.\ Meth.\ A {\bf 593}, 448 (2008).
  
\bibitem{Ohsumi:2002ah} 
  H.~Ohsumi {\it et al.} [NEMO Collaboration],
  Nucl.\ Instrum.\ Meth.\ A {\bf 482}, 832 (2002).
  
   
\bibitem{NLSM} 
 V. Chazal {\it et al.}, 
 Astroparticle Physics, Elsevier, 9, pp.163-172 (1998).

\bibitem{Mei:2005gm} 
  D.~Mei and A.~Hime,
  Phys.\ Rev.\ D {\bf 73}, 053004 (2006).
  
\bibitem{EXO200::2015wtc} 
  J.~B.~Albert {\it et al.} [EXO-200 Collaboration],
  JCAP {\bf 1604}, no. 04, 029 (2016).
  
\bibitem{Obara:2017ndb} 
  S.~Obara [KamLAND-Zen Collaboration],
  Nucl.\ Instrum.\ Meth.\ A {\bf 845}, 410 (2017).
  
\bibitem{Arnaboldi:2002du} 
  C.~Arnaboldi {\it et al.} [CUORE Collaboration],
  Nucl.\ Instrum.\ Meth.\ A {\bf 518}, 775 (2004).
  
\bibitem{Fard:2015pla} 
  A.~D.~Fard {\it et al.},
  AIP Conf.\ Proc.\  {\bf 1672}, 070003 (2015).
  
\bibitem{Ohsumi:2008zz} 
  H.~Ohsumi [NEMO and SuperNEMO Collaborations],
  J.\ Phys.\ Conf.\ Ser.\  {\bf 120}, 052054 (2008).
  
  \bibitem{LS}
http://neutron.physics.ucsb.edu/docs/scintillation$\_$presentation$\_$info.pdf

\bibitem{Yang:2017ybj} 
  H.~B.~Yang {\it et al.},
  arXiv:1703.01867 [physics.ins-det].

\bibitem{Adam:2007ex} 
  T.~Adam {\it et al.},
  Nucl.\ Instrum.\ Meth.\ A {\bf 577}, 523 (2007).

\bibitem{SG}
 https://www.crystals.saint-gobain.com/sites/imdf.crystals.com/files/documents/sgc-scintillation-fiber$\_$0.pdf
%
 
 
\bibitem{Bol:1997} 
  A.~Bolotnikov and B.~Ramsey,
  Nucl.\ Instrum.\ Meth.\ A {\bf 396}, 360 (1997).

\bibitem{Giganon:2017isb} 
  A.~Giganon {\it et al.},
  arXiv:1707.09254 [physics.ins-det].
 
\end{thebibliography}
\end{document}